\documentclass[12pt]{article}
\usepackage{epsfig}
\newcommand{\qq}{${\overline q}q$~}
\newcommand{\gamgam}{$\gamma\gamma$~}
\newcommand{\ggpp}{$\gamma\gamma\to\pi\pi$~}
\newcommand{\ggppch}{$\gamma\gamma\to\pi^+\pi^-$~}
\newcommand{\ggppn}{$\gamma\gamma\to\pi^0\pi^0$~}
\newcommand{\pp}{$\pi\pi$~}
\newcommand{\kk}{${\overline K}K$~}
\newcommand{\pppp}{\pi\pi\to\pi\pi}
\newcommand{\ppkk}{\pi\pi\to{\overline K}K}

\newcommand{\elel}{$e^+e^-$~}
\newcommand{\be}{\begin{equation}}
\newcommand{\ee}{\end{equation}}

\newcommand{\eea}{\end{eqnarray}}
\newcommand{\cost}{\cos\,\theta^*}
\newcommand{\fl}{{\cal F}^I_{J,\lambda}}

\newcommand{\tl}{{\cal T}^I_{J}}
\begin{document}

\large
\hfill\vbox{\hbox{DCPT/08/20}
              \hbox{IPPP/08/10}}
\nopagebreak
\vspace{1.2cm}
\begin{center}
\large{\bf AMPLITUDE ANALYSIS of HIGH STATISTICS RESULTS\\ on $\gamma\gamma\to\pi^+\pi^-$\\ and the TWO PHOTON WIDTH of ISOSCALAR STATES}
\vspace{10mm}

\large{M.R. PENNINGTON}

\vspace{4mm}
{\it Institute for Particle Physics Phenomenology, \\Durham University,
Durham DH1 3LE, United Kingdom\\
m.r.pennington@durham.ac.uk}

\vspace{6mm} 
\large {T. MORI}

\vspace{4mm}

{\it Department of Physics, Nagoya University, Nagoya, Japan}

\vspace{6mm} 
\large {S. UEHARA}

\vspace{4mm}

{\it High Energy Research Organisation (KEK), Tsukuba, Japan}
\vspace{6mm}

\large{Y. WATANABE}

\vspace{4mm}
 
{\it Faculty of Engineering, Kanagawa University, Yokohama, Japan}
\end{center}

\vspace{7mm}
\small

\begin{abstract}

We perform an Amplitude Analysis of the world published data on $\gamma\gamma\to\pi^+\pi^-$ and $\pi^0\pi^0$. These are dominated in statistics by the recently published results from Belle on the charged pion channel. Nevertheless, having only limited angular information, a range of solutions remain possible. We present two 
solutions with $\Gamma(f_0(980)\to\gamma\gamma) = 0.42$ and $0.10$ keV, and
$\Gamma(f_2(1270)\to\gamma\gamma) = 3.14 \pm 0.20$ and $3.82 \pm 0.30$ keV, respectively: the former being the solution favoured by $\chi^2$, the latter at the edge of acceptability.
Models of the structure of the $f_0(980)$ predict two photon widths to be between 0.2 and 0.6 keV, depending on its composition as mainly ${\overline K}K$, ${\overline s}s$ or ${\overline{qq}}qq$. Presently available data cannot yet distinguish unambiguously between these predictions. However, we show how forthcoming results on $\gamma\gamma \to\pi^0\pi^0$ can not only discriminate between, but also refine,  these classes of partial wave solutions.

\end{abstract}


\normalsize
\baselineskip=5.5mm
\parskip=2.mm
\newpage
\section{Introduction}

Two photon processes provide one of the cleanest probes of hadron structure and strong dynamics~\cite{brodsky}. They are an almost unique way of measuring the charges carried by the constituents of any hadronic state. For neutral mesons built of quarks, a state $R$ has two photon couplings given by 
\be
\Gamma(R\to\gamma\gamma)\;=\;\mid\,\sum_q\, e_q^2\,\Pi_R\,\mid^2\quad , 
\ee
where the sum is over all the contributing quarks and in the non-relativistic quark model $\Pi_R$ is the probability that the constituent quarks annihilate. For heavy quarks
 this is proportional to the square of the wave-function at the origin. For light quarks there may be substantial corrections, estimated in Ref.~\cite{rel-corr}
and reviewed in Ref.~\cite{mpreviews}.
\begin{figure}[h]
\centerline{\psfig{file=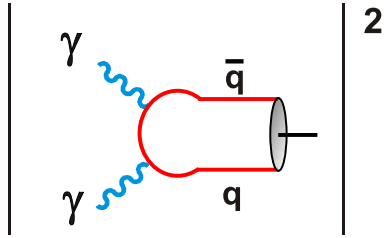,width=5.cm}}
\vspace*{2pt}
\caption{\leftskip 1cm\rightskip 1cm{Two photon decay rate of a meson in a quark picture is the modulus squared of the amplitude for $\,\gamma\gamma\,$ to produce a $\,{\overline q}q\,$ pair and for these to bind by strong coupling dynamics to form the hadron. }} 
\end{figure}

 Consequently,
the two photon production of $\pi^+\pi^-$, for example, can be used to study the nature of all the states that couple to the $\pi\pi$ system.  Remembering that Bose symmetry requires that  odd $J$ does not couple, we can access those states for which $I$ and $J$ are even. Among the most prominent of these is the $f_2(1270)$. Lying underneath this are the \lq\lq enigmatic'' scalars in the $I=J=0$ channel. Very many models have been proposed in the literature for their classification whether ${\overline q}q$, ${\overline {qq}}qq$ or glueball --- see for instance~\cite{mpreviews,klempt,mp-menu} and the references therein. The two photon couplings of these scalars are one of the clearest guides to their compositions. However,
the $\sigma$, $f_0(980)$, and the many higher $f_0$'s, all overlap with each other. This means that one is not looking to fit data with sums of  Breit-Wigner forms and simply extract couplings to \gamgam of each resonance, but rather we have to perform a complete Amplitude Analyses to separate all the contributing spin components. Only then can we determine the couplings of the meson states
by continuing these amplitudes to each resonance pole in turn.

Data on the $\gamma\gamma\to\pi^+\pi^-$ channel~\cite{datareview} have existed for many years, almost from the start of $e^+e^-$ colliders. Some of the earliest results, like that from DM2~\cite{dm2}, were merely event distributions, but with data from
Mark II~\cite{boyer} at PEP twenty years ago followed by CELLO~\cite{harjes,behrend} at PETRA, normalised cross-sections for charged pion production became available. With the Crystal Ball detector first running at SLAC and then moving to DESY, normalised cross-sections for $\pi^0\pi^0$ production in two photon collisions  also resulted~\cite{cb88,cb92}.
With these an Amplitude Analysis using a methodology that provides the nearest one can presently get to
a model-independent partial wave separation of $\gamma\gamma\to\pi^+\pi^-$,
$\pi^0\pi^0$ cross-sections became feasible~\cite{shoresh,morgam1,morgam2}. This resulted in two distinct classes of solutions~\cite{boglione}, distinguished by whether the $f_0(980)$ appears as a ``peak'' or a ``dip''. Data in wide energy bins made both these possible. In the ``peak'' solutions the two photon width of  the
$f_0(980)$ was found to cover a range from 130 to 360 eV~\cite{boglione}, as cited in the PDG tables~\cite{pdg}. 

We now have data on charged pion production with a hundred times the statistics from Belle~\cite{mori,abe} allowing bins of 5 MeV width for both integrated and differential cross-sections. These reveal a clear
peak in the 950-980 MeV region. Fitting the Belle integrated cross-section with model forms~\cite{abe} gives a two photon width for the $f_0(980)$ of $\Gamma(f_0)\,=\,205 ^{+95+147}_{-83-117}$ eV. With the dramatically increased statistics of these results it is natural to consider how these improve the Amplitude Analysis, when the full data set (integrated and differential) are folded with the earlier world data. That is the aim of this work. 

We will find that the range of possible solutions is now more readily specified than in earlier work~\cite{morgam2,boglione}. 
With the large dataset from Belle, a two photon width for the $f_0(980)$ of 415 eV is favoured. However, solutions with a  width 
from 100 to 540 eV are all acceptable. That such a wide range is still possible is partly because of the large systematic uncertainties in the Belle $\pi^+\pi^-$ angular distributions. However, these solutions can be resolved by precision $\pi^0\pi^0$ data particularly in key energy regions.
To understand how we arrive at these conclusions, 
we outline the methodology in Sect.~2, present  our solutions for $\gamma\gamma$
amplitudes solutions in Sect.~3 with plots of data fitted, and then in Sect.~4 discuss the results for resonance two photon widths, the implications for hadronic structure and give our conclusions.

\section{Methodology}

For a process, where one measures all observables and data cover the complete $4\pi$ angular range, such an Amplitude Analysis would be reasonably straightforward. However, here the photons are unpolarised and  we have no information on the initial spin. Moreover, the detection of the charged final state in an environment of \elel means that pions are only identifiable in the centre of the angular range in the \gamgam rest system, typically $|\cos \theta^*|\le 0.6$. Consequently, one has large areas of the reaction plane with no direct experimental information. The effect of this is readily illustrated by an example.

\begin{figure}[t]
\centerline{\psfig{file=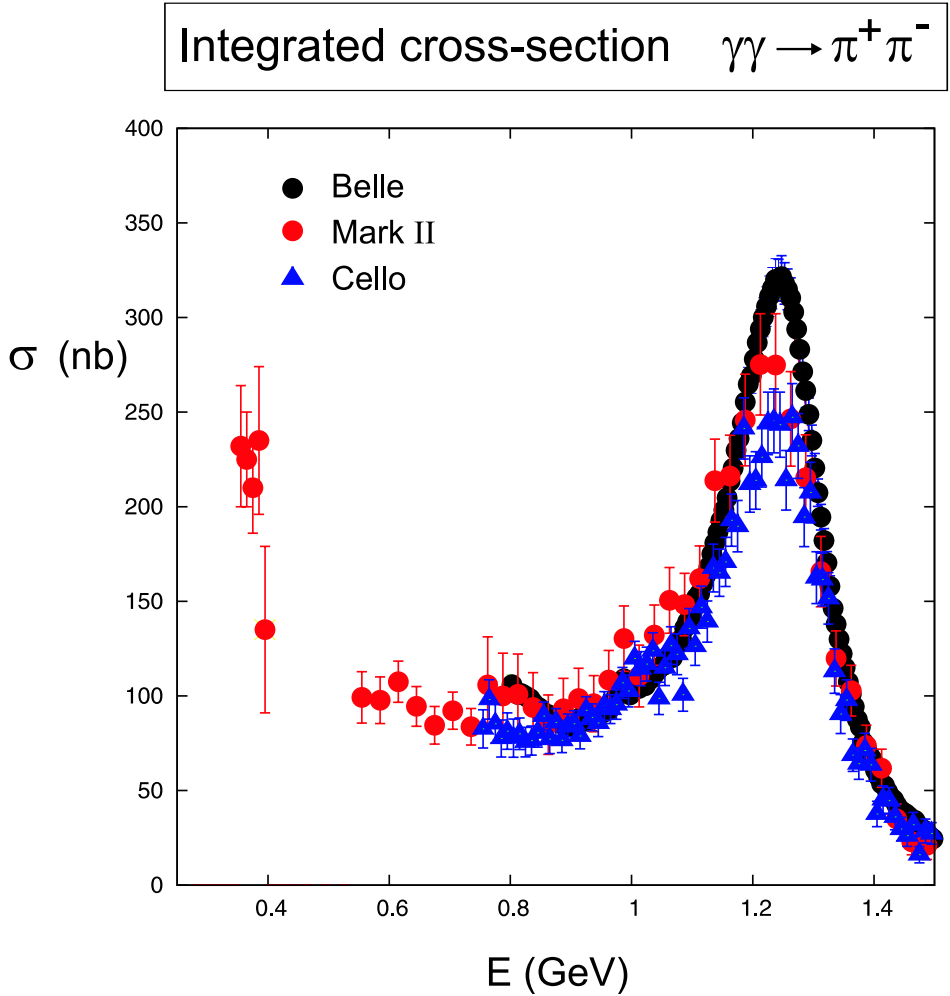,width=12.5cm}}
\vspace*{2pt}
\caption{\leftskip 8mm\rightskip 8mm{Comparison of the cross-section results for $\,\gamma\gamma\to\pi^+\pi^-\,$ from Mark~II~\cite{boyer}, Cello~\cite{harjes} and Belle~\cite{abe}. In each case the cross-section is integrated over $|\cos \theta^*|\,\le\,0.6$.}}
\end{figure}
In Fig.~2 we show a compilation of the world data on \ggppch from PEP, PETRA and BELLE~\cite{boyer,harjes,abe}. All are integrated over the range of $|\cos\theta^*| \le 0.6$.
The dominant feature of the
\ggpp cross-section below 1.5 GeV is the $f_2(1270)$ peak. This is a well-known spin-2 \qq resonance with $I=0$. It can be formed in two photon reactions with either helicity 0 or~2. The non-relativistic quark model~\cite{schrempp} suggests pure helicity~2. Relativistic corrections have been found to be 
small~\cite{rel-corr}. What we observe is a large peak in the \ggpp cross-section.
Let us assume we have picked out the total $I=0$ $J=2$ component. The angular distribution will then, in principle, tell us how much has helicity $0$ and how much helicity $2$. In Fig.~3, the upper three plots show this distribution in a bin centred on 1270 MeV with different combinations of helicity 0 and 2. If we had data covering the complete angular range up to $|\cos \theta^*| \,=\,1$, the separation of these two helicity components would be relatively easy. However, if we now fold in the angular acceptance that cuts off at $|\cos \theta^*|\, =\,0.6$, we arrive at the lower 3 plots. We see that these are virtually indistinguishable. To solve this problem, we need two things: firstly data of very high precision in the angular range covered. Secondly we need additional theoretical information to make up for the lack of complete angular coverage.
Though the published results on $\pi^+\pi^-$ production from Belle~\cite{abe} have sufficient statistical precision, systematic uncertainties mean that a large variation in the ratio of helicity~2 to helicity~0  couplings of the $f_2(1270)$, and in the radiative width of the $f_0(980)$, are still possible. As we shall see, precision $\pi^0\pi^0$ angular distributions, when finalised, are likely to resolve these
ambiguities. With the present $\pi^+\pi^-$ data, the ambiguity displayed  in Fig.~3 largely remains. The solutions with the larger two photon width for the $f_0(980)$
have a smaller $D_0$ component, while the smaller $S$-wave have a correspondingly larger $D_0$ contribution. While the quark model for the $f_2(1270)$ would 
favour a small $D_0$ contribution, and the Belle data~\cite{abe} are consistent with this, our  Amplitude Analysis does allow acceptable solutions with a larger $D_0$ component.
 To see what progress can be made  we need to understand the role of analyticity, crossing, unitarity and the low energy theorems of QED and chiral dynamics~\cite{mpdaphne}.
\begin{figure}[t]
\centerline{\psfig{file=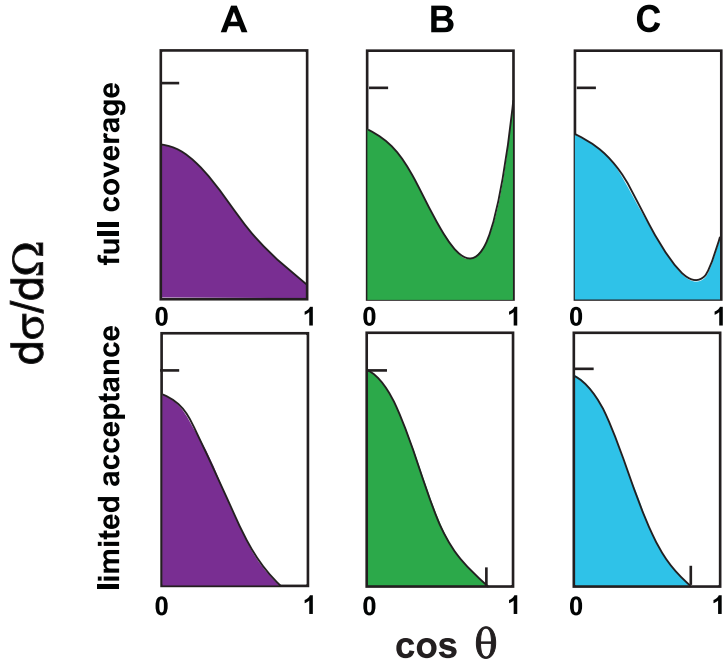,width=9.5cm}}
\vspace*{2pt}
\caption{\leftskip 1cm\rightskip 1cm{The upper plots show 3 examples of the angular distribution in the region of the $f_2(1270)$ with distinctly different ratios of $I=0$ $S\,:\,D_0\,:\,D_2$ components. The lower plots show how these examples are typically modified when the  angular efficiency fades away rapidly beyond $\cos \theta^* =0.6$.}}
\end{figure}

\begin{figure}[p]
\centerline{\psfig{file=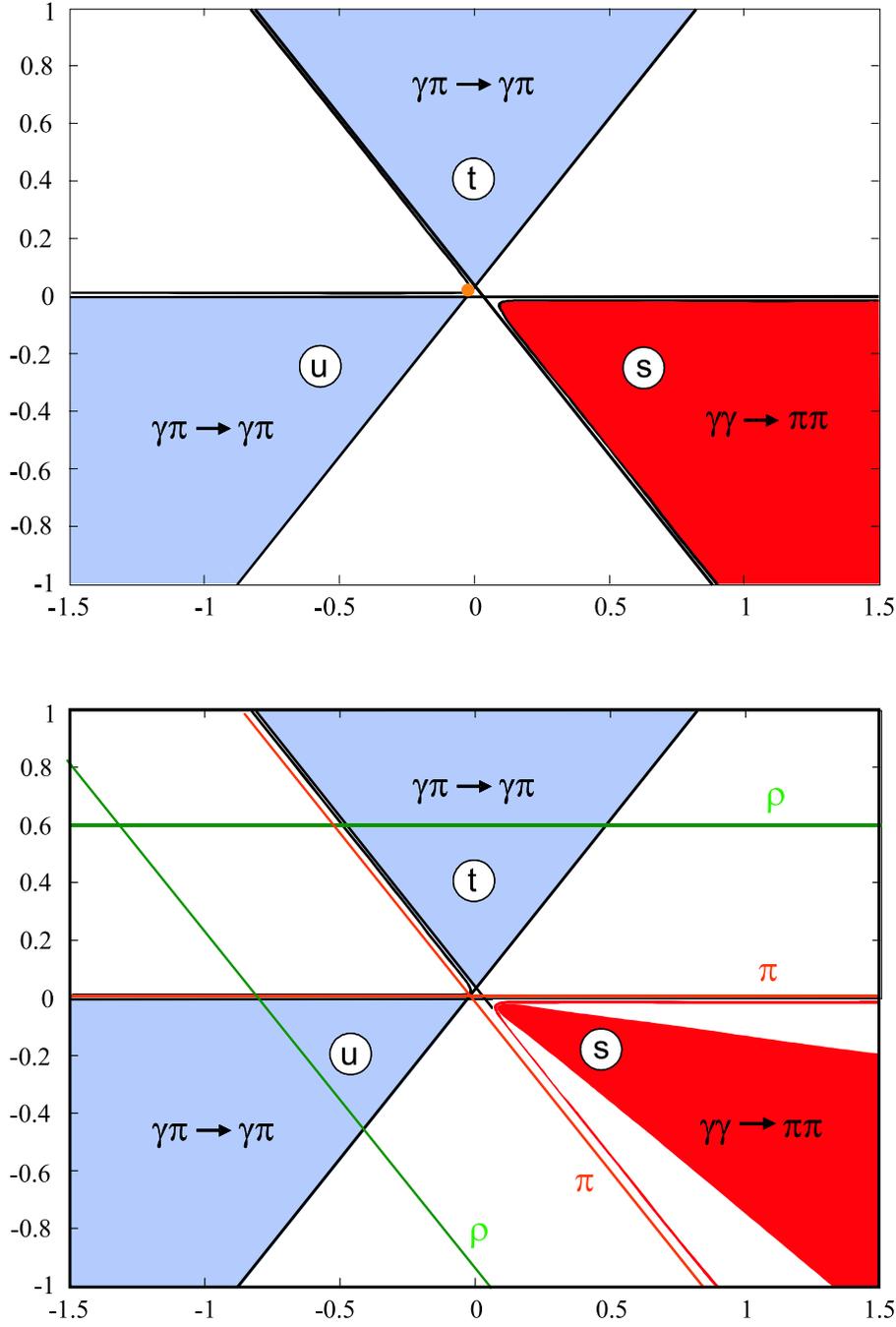,width=12.cm}}
\vspace*{2pt}
\caption{\leftskip 1cm\rightskip 1cm{Mandelstam plane with $\gamma\gamma\to\pi^+\pi^-$ in the $s$-channel and Compton scattering off a charged pion in the $t$ and $u$-channels. The (orange) dot marks the common threshold for the two Compton processes, where the Thomson limit applies. In the lower plot, drawn to scale, the shaded region of the $s$-channel delineates the physical region with $-0.6\, \le\,\cos \theta^*\,\le\, 0.6$. The position of the $t$ and $u$-channel pion poles indicating how very close they are to the $s$-channel physical region. The position of the $\rho$ poles are shown, showing how far away the next most important crossed channel exchange are. The figures on the axes are in GeV$^2$.  }}
\end{figure}

It is convenient to consider the Mandelstam plane for the process \ggpp in the $s$-channel. This is shown in the upper plot of Fig.~4.
The crossed-channels are, of course, $\gamma\pi\to\gamma\pi$. At the threshold for this reaction, marked by the dot in Fig.~4, the photon has no energy and so has infinite wavelength. The probing photon then just sees the charge on the pions. Indeed, it is this Thomson limit of Compton scattering that defines the charge on the pion. At this point there is a low energy theorem due to Low~\cite{Low} that requires the Compton amplitude to be equal to its Born term. This is given by one pion exchange in the $t$ and $u$-channels and a contact interaction to preserve gauge invariance.
Thus at one point in this Mandelstam plane, we know the \ggppch amplitude exactly. Abarbanel and Goldberger~\cite{goldberger} showed this limit is approached in a smooth way so that the amplitude in the neighbourhood of the Thomson limit is still controlled by the one pion exchange amplitude.
This is not really surprising since the pion mass is so very small. The pion poles at $t$ and $u\simeq 0.02$ GeV$^2$ are so much closer to the threshold than the contributions from any other exchange, such as that of the $\rho$ with poles at $t$ and $u\simeq$ 0.5 GeV$^2$.
The position of these poles are drawn to scale in Fig.~4. In the lower plot we see the effect the experimental cut on $\cost$ has on where we have data on \ggppch in the $s$-channel physical region. The nearness of the pion pole to the forward and backward directions provides the key to how we can use the calculability of final state interactions, to make up for our lack of knowledge of the
cross-section for $\cost > 0.6$ --- at least at low energies, as we discuss briefly below.

\baselineskip=5.5mm

One important fact is that since the Born term that controls low energy \ggpp has $I=1$ exchange in the $t$ and $u$-channels, this involves both $I=0$ and 2 in the $s$-channel. Usually in reactions in which hadrons are produced in the final state, $I=2$ amplitudes are generally much smaller than those with $I=0$ (or 1). This is because there are no established resonances with $I=2$: indeed these are of course {\it exotic} in the quark model. Here in \ggpp at least at low energy, $I=2$ amplitudes are equally important, otherwise the Thomson limit could not be
maintained. This means that the reaction \ggppch cannot be analysed in terms of amplitudes with definite isospin on its own, despite earlier attempts to do so.
One must also have comparable information on the \ggppn channel too. Fortunately this comes from an earlier experiment by Crystal Ball
at DESY~\cite{cb88} with results limited to $|\cost| \le 0.8$. Our Amplitude Analysis treats both reactions simultaneously. The low energy theorem for \ggppn is trivial, namely the amplitude is zero --- neutral pions having no charge. Away from the Thomson limit, the chiral nature of pion interactions limits the scope for variation, as detailed in Refs.~\cite{morgam1,morgam2,mpdaphne,mpreviews}.

To proceed we first separate the amplitude for \ggpp into its helicity components and then each of these into their partial waves. 
The unpolarized cross-section for dipion production by two
real photons is given by the contribution of two helicity amplitudes
\begin{figure}[h]
\vspace{3mm}
\centerline{\psfig{file=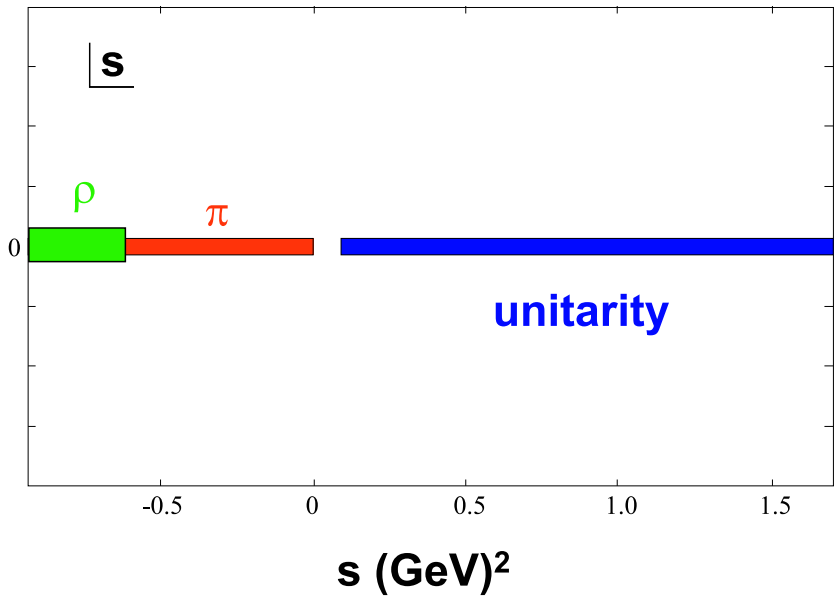,width=11.cm}}
\vspace*{2pt}
\caption{\leftskip 1cm\rightskip 1cm{Cuts in the complex $s$-plane of the partial wave amplitudes for $\gamma\gamma\to\pi\pi$, drawn to scale, illustrating how the crossed channel pion exchange controls the nearby part of the left hand cut.}}
\end{figure}
\noindent $M_{++}$ and $M_{+-}$ (the subscripts label the helicities of the
incoming photons) : 
\begin{equation}
\frac{d \sigma}{d \Omega} = \frac{1}{128 \pi ^2 s} \sqrt{1- 4m^2 _{\pi} /s}
\; \left[ \; |M_{++}|^2 + \, |M_{+-}|^2 \, \right] .
\end{equation}
These two helicity amplitudes can be decomposed into partial waves as
\begin{equation}
M_{++}(s,\theta,\phi) = e^2 \sqrt{16 \pi} \sum _{J \geq 0} F_{J0}(s) \;
Y_{J0}(\theta, \phi) \; ,
\end{equation}
\begin{equation}
M_{+-}(s,\theta,\phi) = e^2 \sqrt{16 \pi} \sum _{J \geq 2} F_{J2}(s) \;
Y_{J2}(\theta, \phi) \; .
\end{equation}
The partial waves $F_{J \lambda} \, (\lambda=0,2)$ are the quantities we want 
to determine.

Each partial wave $\fl$ is then an analytic function of $s$ the square of the centre-of-mass energy $E$. Each has a right hand cut starting at the two pion threshold. Importantly, they also have a left hand cut
controlled by crossed-channel exchanges --- Fig.~5. As we have emphasised,
the part of this nearest to the direct channel is controlled by pion exchange as we have discussed. Corrections come from $\rho$, $\omega$, $a_1, \cdots$ exchange, but these have a small
 effect on the \ggpp partial waves below 500 MeV. What is more important are  $\pi\pi$ final state interations, especially in the $I=J=0$ channel. Dispersion relations~\cite{morgam0} provide the vehicle for translating knowledge of this nearby left hand cut, the low energy theorems and experimental information on meson-meson scattering into precise information about the behaviour of the \ggpp partial waves at low energy. This is described in Refs.~\cite{morgam1,mpdaphne,mp-prl}. The recent work of Oller {\it et al.}~\cite{oller} confirms our calculation in the near threshold domain. A key aspect of this calculability of the final state interactions is that the
discontinuity across the right hand cut is constrained by unitarity, as shown in Fig.~6. The sum on the right hand side is over all allowed intermediate states: allowed by quantum numbers and kinematics. At low energy only $\pi\pi$ contributes and the well-known Watson's theorem~\cite{watson} applies. Since the $4\pi$ channel is known to be weak, $\overline{K}K$ is the next most important contribution. Inclusion of just this channel allows our analysis to be extended up to
1400 MeV, then multi-meson states become critical as $\rho\rho$ threshold is approached.
\begin{figure}[h]
\vspace{3mm}
\centerline{\psfig{file=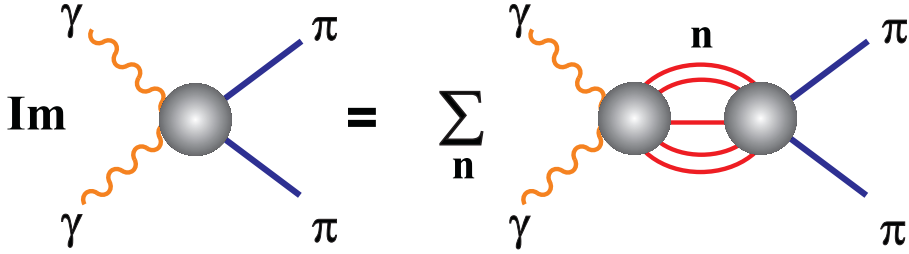,width=12.cm}}
\vspace*{2pt}
\caption{\leftskip 1cm\rightskip 1cm{Unitarity relation for each partial wave of $\gamma\gamma\to\pi\pi$.}}
\end{figure}

If each partial wave for the reaction $\pi\pi\to$ hadronic final~state $H_n$ with a given isospin is denoted  by $\tl$, then coupled channel unitarity
is solved by requiring for $s > 4m_{\pi}^2$
\be
\fl(s)\;=\;\alpha^I_{J\lambda}(s)\,\cdot\,\tl(s) \quad ,
\ee 
where $\alpha(s)$ is an $n$-component vector of {\bf real} coupling functions representing the coupling of \gamgam to channel $H_n$.
 This reflects the fact that the right hand cut structure of the \ggpp amplitude is that of the hadronic processes  with \pp final states. The functions $\alpha(s)$ have no right hand cut.\footnote{Very recently van Beveren and Rupp~\cite{beveren,wilson} have proposed that there is merit in writing these real functions in terms of particular complex functions. Though this can indeed be done, it introduces a right hand cut structure into these functions. The simplicity of our coupling functions $\alpha(s)$ is that they are real and having only left hand cuts are smooth for $s> 4m_{\pi}^2$.} In general, there would also be a helicity label on the hadronic amplitudes too, but we will limit our consideration to spinless intermediate states like \pp and \kk. Thus for the most part we will consider the energy region up to about 1400 MeV, for which a two channel representation is adequate, then more explicitly, we have
\be
 \fl(s;\gamma\gamma\to\pi\pi)\;=\;{\alpha_1}^I_{J\lambda}(s)\,\hat\tl(s;\pppp)\,+\,{\alpha_2}^I_{J\lambda}(s)\,\hat\tl(s;\ppkk)\; .
\ee
This is rendered useful by knowledge of these two meson-meson scattering amplitudes determined by combining experimental information, with unitarity and analyticity constraints~\cite{bpamps}. This provides a representation of the $\pppp$ and $\ppkk$ amplitudes from \pp threshold to 1400 MeV. The \lq hat' indicates \lq\lq reduced'' amplitudes~\cite{AMP}, in which for the $S$-wave alone the process-dependent Adler zeroes in the $\pi\pi\to\pi\pi$ and ${\overline K}K$ amplitudes have been divided out.{\footnote{The $I=J=0$ hadronic amplitudes used~\cite{bpamps} for the two-meson channels $\pi\pi$ and ${\overline K}K$ in the initial and final states have no zero in their determinant, and so the additional complications discussed in Refs.~\cite{AMP,boglione} to parametrise the coupling functions is not required.}
 Then Eq.~(6) ensures that two body unitarity is respected for each $\gamma\gamma$ partial wave amplitude. The real functions
$\alpha_n(s)$ are the parameters in terms of which we fit the world two photon
data. As already discussed, these are strongly constrained (indeed even precisely known) down below 600 MeV by the Born term modified by calculable final state interactions. This serves to anchor our Amplitude Analysis at low energy.
The use of the parametrization in Eq.~(6) imposes continuity on our partial wave solutions. This helps to limit the ambiguity illustrated in Fig.~3 by relating what happens in one energy bin to that in its neighbours. In this way we perform an energy dependent amplitude analysis. With the latest data being in 5 MeV bins this is particularly appropriate. This allows us to vary just 16-22 parameters to fit more than 2200 data points as we discuss next. 

\newpage
\section{Amplitudes determined: solutions A and B}

\baselineskip=5.6mm

Armed with the methodology of Sect.~2 and the representation of Eq.~(6), we can  fit the complete set of experimental cross-sections, integrated and differential, listed in Table~1 by expressing the functions $\alpha_1(s), \alpha_2(s)$, in terms of smooth functions exactly as in Refs.~\cite{morgam2,boglione}. In Table~1 we see that the number of data-points is far greater 
for the reaction $\gamma \gamma \to \pi ^+ \pi ^-$ than for
 $\gamma \gamma \to \pi ^0 \pi ^0$. Since an accurate separation of 
 the $I=0$ component requires both are accurately described, we give
  different weight factors to each dataset. We choose these so that the
   Mark II, CELLO and Belle data have roughly the same number of weighted data, 
   while the weight assigned to the Crystal Ball data approximately equals 
   the weighted sum of the $\pi^+\pi^-$ data. 
Nevertheless, good agreement is not easy to achieve.

\noindent The analysis program 
 works by integrating the amplitudes
over the appropriate bin in energy and angle for each data-point. 
It  does not just use the
central values. This is to allow for any strong local variation of the amplitudes,
particularly near $K{\overline K}$ threshold. In the many figures displaying the
solutions we find, this should be borne in mind.  Where the energy bins are sizeable
(as with Crystal Ball) histograms are plotted (see Fig.~8).
 Where the energy bins are fine (as with Mark II data in
10 MeV steps and Belle data in 5 MeV bins), 
the fits are shown more appropriately as continuous lines joining the
bin centres, as in Figs.~7, 9, 10, 11 and 12. 
However throughout, the fits are histograms of appropriate bin width.
\begin{table}[h]
\vspace{0.2cm}
\caption{\leftskip 0.5cm\rightskip 0.5cm{Number of data in each experiment
 below 1.44 GeV. Mark II results are from Boyer {\it et al.}~\cite{boyer},
 Crystal Ball from Marsiske {\it et al.} (CB88)~\cite{cb88} and Bienlein 
 {\it et al.} (CB92)~\cite{cb92}, 
 CELLO from Harjes~\cite{harjes} 
 and Behrend {\it et al.}~\cite{behrend}, and Belle from Mori {\it et al.}~\cite{abe}. }}
\begin{center}
\begin{tabular}{||c|c||c|c||c|c||}
\hline \hline
\rule[-0.4cm]{0cm}{12mm} Experiment & Process & 
 Int. X-sect. & $|\cos \theta |_{max}$ & Ang. distrib. 
 & $|\cos \theta |_{max}$ \\ 
\hline \hline
\rule[-0.75cm]{0cm}{15mm} 
Mark II & $\gamma \gamma \to \pi ^{+} \pi^{-}$ & 81 & 0.6 & 63 & 0.6 \\  
\hline 
\rule[-0.75cm]{0cm}{15mm}   
Cr. Ball & $\gamma \gamma \to \pi ^{0} \pi^{0}$ & 36 & 
\parbox{2.0cm}{0.8~(CB88) \protect \\ 0.7~(CB92)} & 90 & 0.8 \\
\hline
\rule[-0.75cm]{0cm}{15mm}   
CELLO & $\gamma \gamma \to \pi ^{+} \pi^{-}$ & 28 & 0.6 &  
\parbox{2.4cm}{104~(Harjes) \protect \\ 201~(Behrend)} & 0.55 - 0.8 \\  
\hline
\rule[-0.75cm]{0cm}{15mm}
Belle & $\gamma\gamma\to\pi^+\pi^-$ & 128 & 0.6 & 1536 & 0.6\\
\hline \hline 
\end{tabular}
\end{center}
\end{table}

\begin{figure}[t]
\centerline{\psfig{file=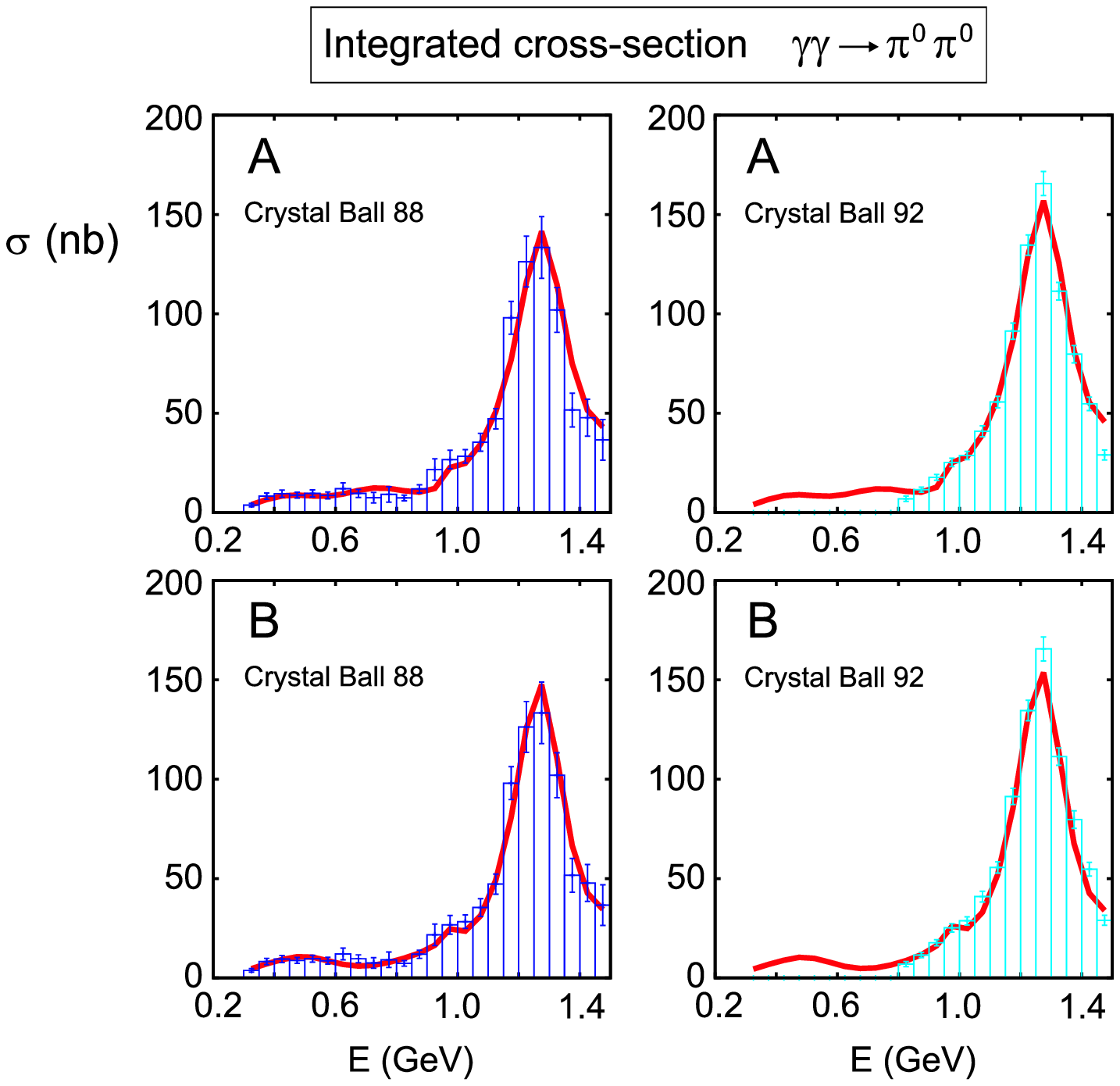,width=13.5cm}}
\vspace*{2pt}
\caption{\leftskip 1cm\rightskip 1cm{Solutions A and B compared with the Crystal Ball results on \ggppn. The 1988 Crystal Ball data~\cite{cb88} are integrated over $|\cos \theta^*|\,\le\,0.8$, while the 1992 data~\cite{cb92} with increased statistics cover $|\cos \theta^*|\,\le\,0.7$.}}
\end{figure}

The solutions are determined by the published statistical errors. In the case of the Belle results, the point-to-point systematic errors are folded into these statistical 
uncertainties. This still leaves the Belle results as the most precise of all the two photon data available, as illustrated in Fig.~2. However, only CELLO have folded their full systematic uncertainties into their published results. For all others there is a global systematic uncertainty that allows all the cross-sections across all energies to be renormalised. For Mark II this is by $11\%$, for Crystal Ball the shift is $11\%$ for the 1988 results and by $3\%$ for their 1992 data. Only for Belle is the energy dependence of the systematics determined. This is typically $12\%$, completely dominating the statistical errors.  We therefore quote the $\chi^2$ in two ways for our solutions.
We give that fixed by the \lq\lq statistical'' errors and in brackets that given by folding in the full global systematic uncertainties in quadrature. In our plots, displaying solutions against the data they fit, the overall systematic errors are {\bf not} included.

The Mark II, Crystal Ball and Belle datasets~\cite{boyer,cb88,cb92,abe} each have a systematic normalization uncertainty. To fit the Amplitudes to the competing datasets, we allow each to shift within these bands. In the Solutions we display, rather than plot the published data shifted, we choose to rescale the fitted amplitudes. Consequently, the solutions displayed in Figs.~7-18 may appear slightly shifted between one dataset and another. This is most evident in Fig.~8, where the near threshold normalization of Mark II differs from that above 450 MeV. Though the Belle data~\cite{abe} have an overall  11-12\% systematic uncertainty, the solutions generally take up no more than a 3\% shift.

In terms of $\chi^2$, Solution A is favoured and provides a very good description of the world data.
While the \lq\lq statistical'' errors give a $\chi^2$ per degree of freedom of 2.9 for this solution, with systematic uncertainties folded in this
reduces to 0.9.
 This solution has a prominent $f_0(980)$ peak in the $I=J=0$ cross-section, and a smaller $D_0$ component throughout the energy region covered. Consequently, this solution predicts a prominent $f_0(980)$ signal in the  $\gamma\gamma\to\pi^0\pi^0$ cross-section. 

When systematic uncertainties are allowed for, Solution A has a probability of 99.9\%. The range of other acceptable solutions is somewhat subjective, largely because of the complex systematics as well as inconsistencies between datasets, as seen in Fig.~2. The criterion we use is not precisely
defined statistically, but based on the goodness of fit to the eye. We have explored the space by requiring varying amounts of $D_0$ to $D_2$ components, as described in Refs.~\cite{morgam2,boglione}. Qualitatively, the solutions are either like Solution A, but with worse $\chi^2$, or closer to what we now describe as Solution B. This is the second solution we display. This solution has an overall $\chi^2$ of 1.2 per degree of freedom (being 3.7 from wholly \lq\lq statistical'' errors). Though this solution only has a probability of $2.10^{-7}$, it provides an acceptable fit to the Belle $\pi^+\pi^-$ data, when systematic uncertainties are folded in. It is the other datasets it mistreats. Solution B
has a much smaller $f_0(980)$ signal in the $I=J=0$ cross-section, while having a larger $I=0$ $D_0$ component. The fit to the distinct peak in the Belle cross-section  between 950 and 1000 MeV for $\gamma\gamma\to\pi^+\pi^-$ in this solution
is just as good as that in the favoured solution. The peak in the $\pi^+\pi^-$ cross-section, integrated over $\cos\theta^*$ from $-0.6$ to $+0.6$, is reproduced by a constructive interference between the $I=0$ $S$ and $D_0$ components.

Figs.~7-10  show how well Solutions A and B fit the integrated cross-sections. $\chi^2$ favours amplitude A, but B is an example at the edge of acceptability.
In all cases the fitting of the Belle $\pi^+\pi^-$ results for the integrated cross-section between 800 and 900 MeV is not good --- see Fig.~10 (and the comparison of data in Fig.~2). This is surely correlated with the strange $\cos \theta^*$ dependence in the Belle results~\cite{abe} shown in Fig.~15 in the same mass region. This is most likely generated by the strong sensitivity to the details of the separation of the large $\mu^+\mu^-$ background in this experiment.

\begin{figure}[p]
\vspace{3.mm}
\centerline{\psfig{file=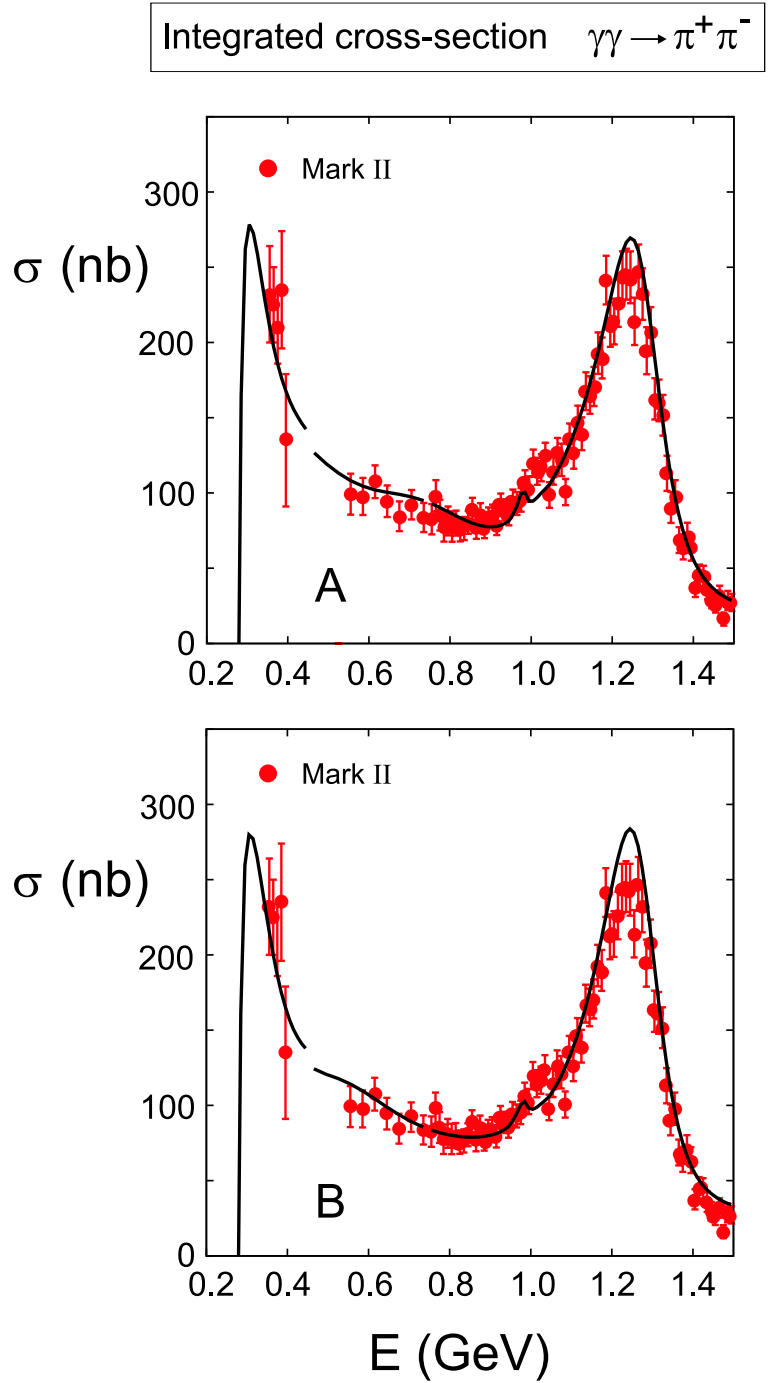,width=10.cm}}
\vspace*{2pt}
\caption{\leftskip 1cm\rightskip 1cm{Solutions A and B compared with the Mark II results~\cite{boyer} on
\ggppch integrated over $|\cos \theta^*|\,\le\,0.6$. The freedom to make a small systematic shift in normalisation of the data above 0.4 GeV has been used to improve the fits. Rather than shift the data, the solutions have been renormalised. This results in the discontinuity in the solid curves at 0.45 GeV}}
\end{figure}

\begin{figure}[p]
\centerline{\psfig{file=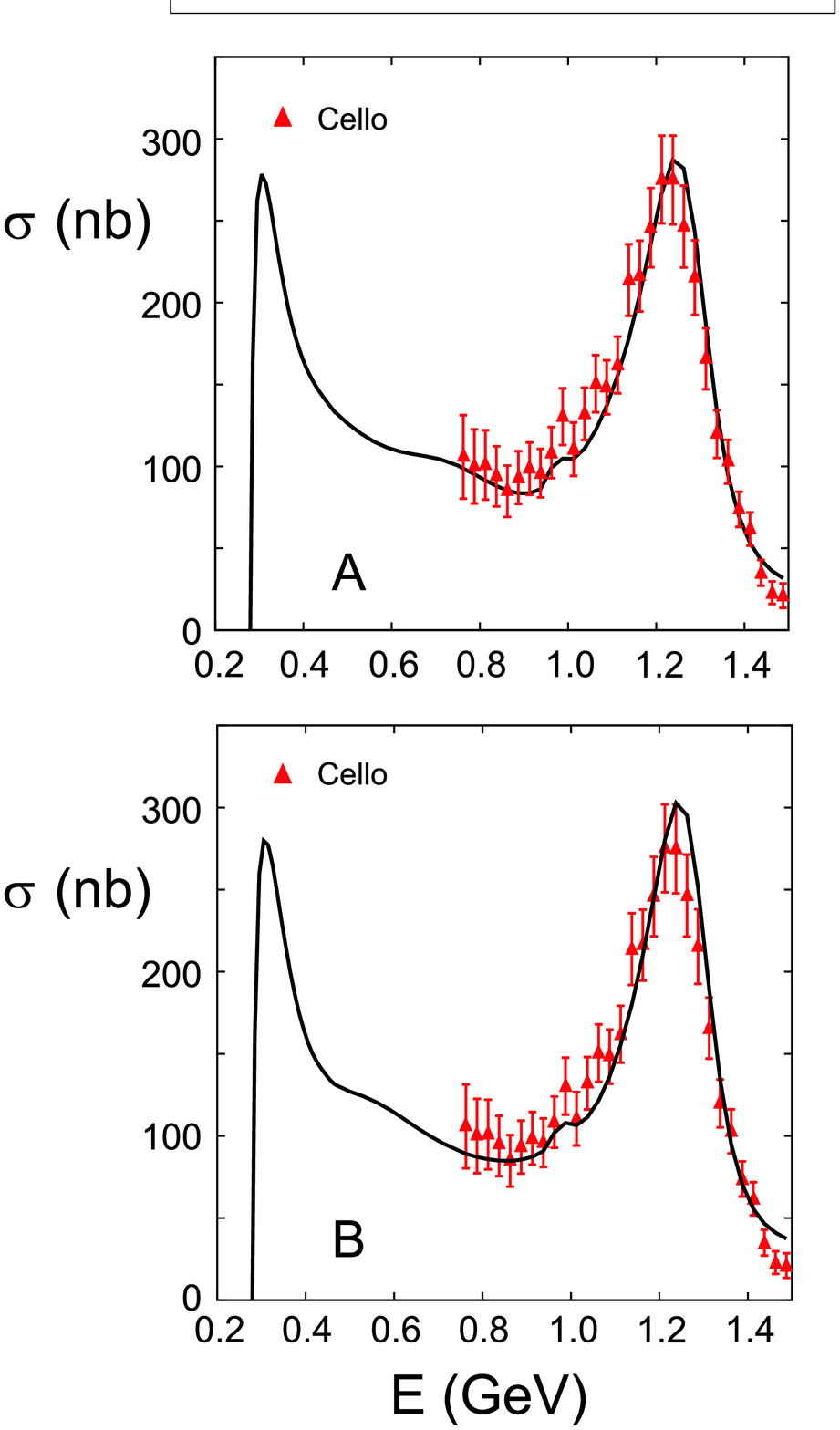,width=10.cm}}
\vspace*{2pt}
\caption{\leftskip 1cm\rightskip 1cm{Solutions A and B compared with the CELLO results~\cite{harjes,behrend} on \ggppch integrated over $|\cos \theta^*|\,\le\,0.6$.}}
\end{figure}

\begin{figure}[p]
\centerline{\psfig{file=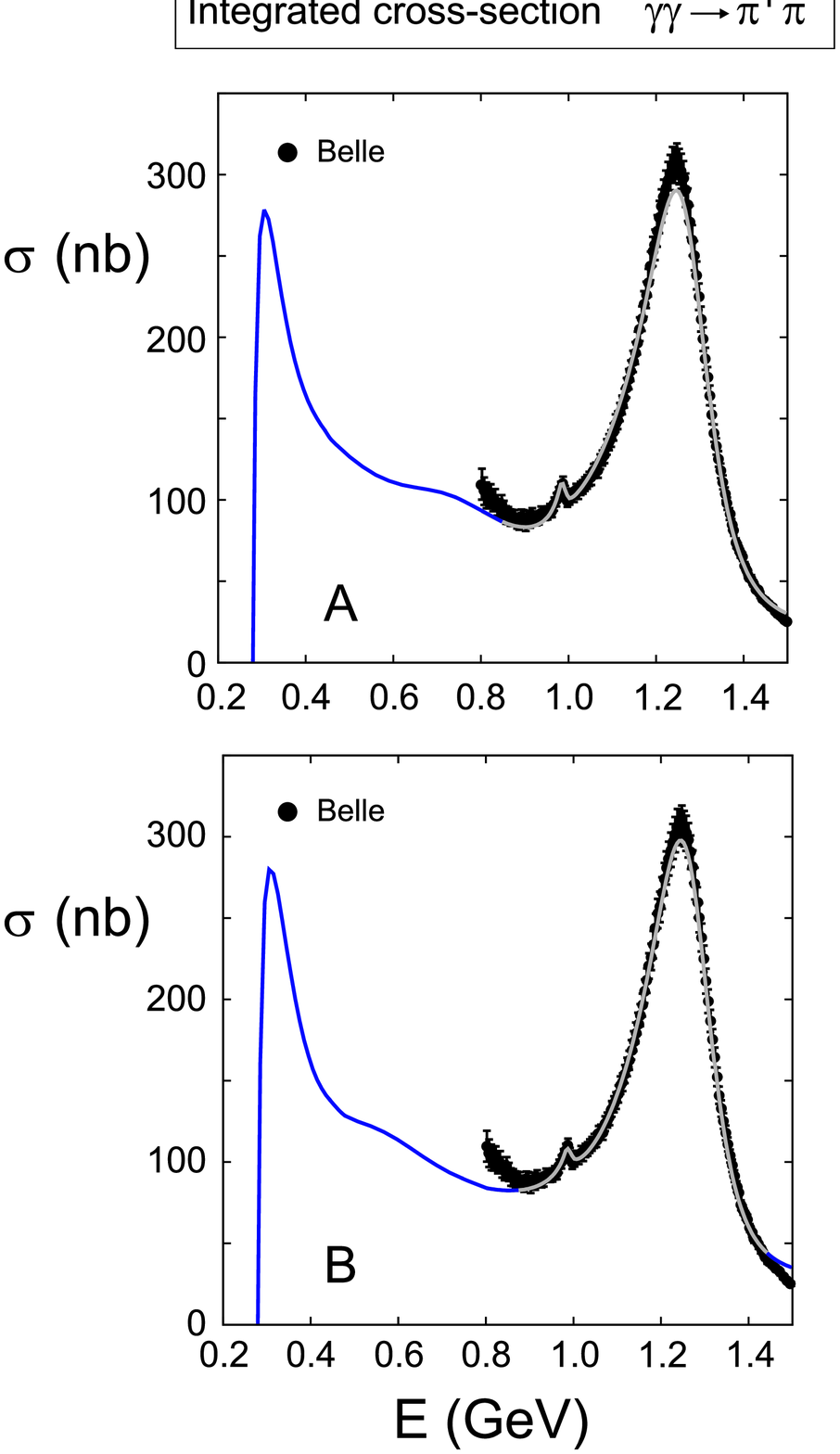,width=10.cm}}
\vspace*{2pt}
\caption{\leftskip 1cm\rightskip 1cm{Solutions A and B compared with the Belle results~\cite{abe} on \ggppch integrated over $|\cos \theta^*|\,\le\,0.6$.}}
\end{figure}

\begin{figure}[h]
\centerline{\psfig{file=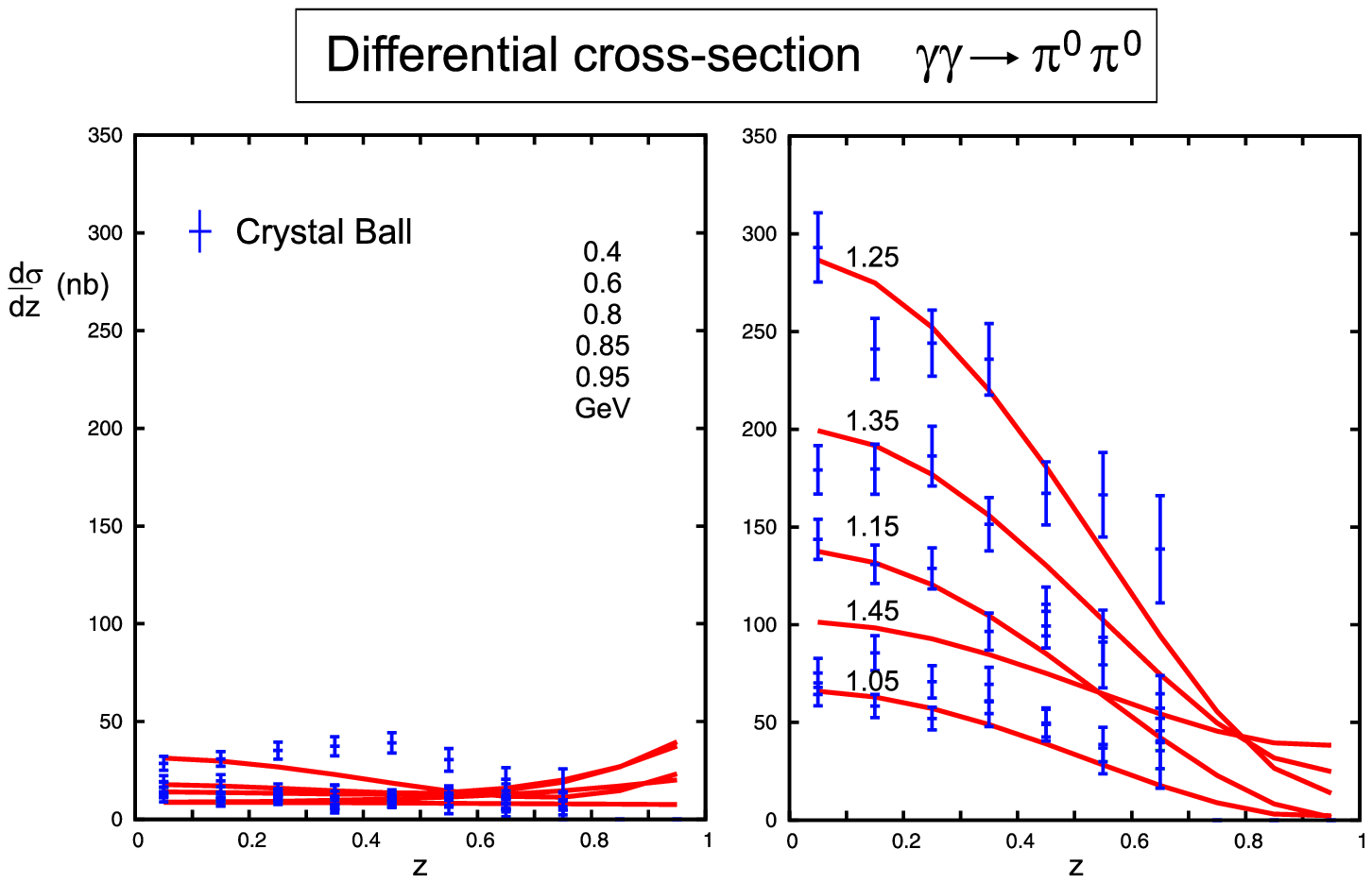,width=12.5cm}}
\vspace*{2pt}
\caption{\leftskip 1cm\rightskip 1cm{Differential cross-section for $\,\gamma\gamma\to\pi^0\pi^0\,$ from the Crystal Ball experiment~\cite{cb88,cb92} compared with Solution A from the Amplitude Analysis. The numbers give the central energy in GeV of each angular distribution listed in order of the size of the  cross-section at $z=0$, where
$z\,=\,\cos\,\theta^*$.}}
\vspace{7mm}
\end{figure}
\begin{figure}[h]
\centerline{\psfig{file=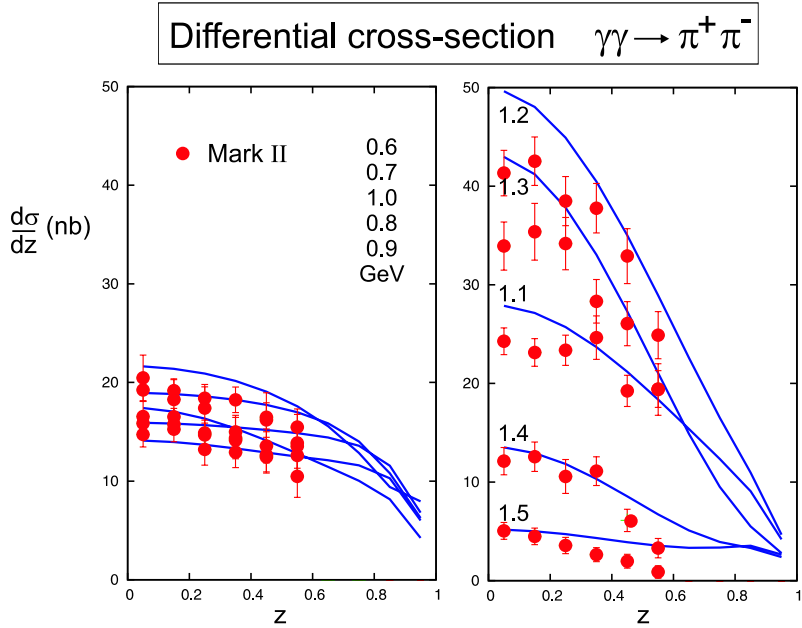,width=10.5cm}}
\vspace*{2pt}
\caption{\leftskip 1cm\rightskip 1cm{Differential cross-section for $\,\gamma\gamma\to\pi^+\pi^-\,$ from the Mark II experiment~\cite{boyer} compared with Solution A from the Amplitude Analysis. The numbers give the central energy in GeV of each angular distribution listed in order of the cross-section at $z=0$, where
$z\,=\,\cos\,\theta^*$. The data are normalized so that the  integrated cross-section is just a sum of the differential cross-sections in each angular bin.}}
\end{figure}


\begin{figure}[p]
\centerline{\psfig{file=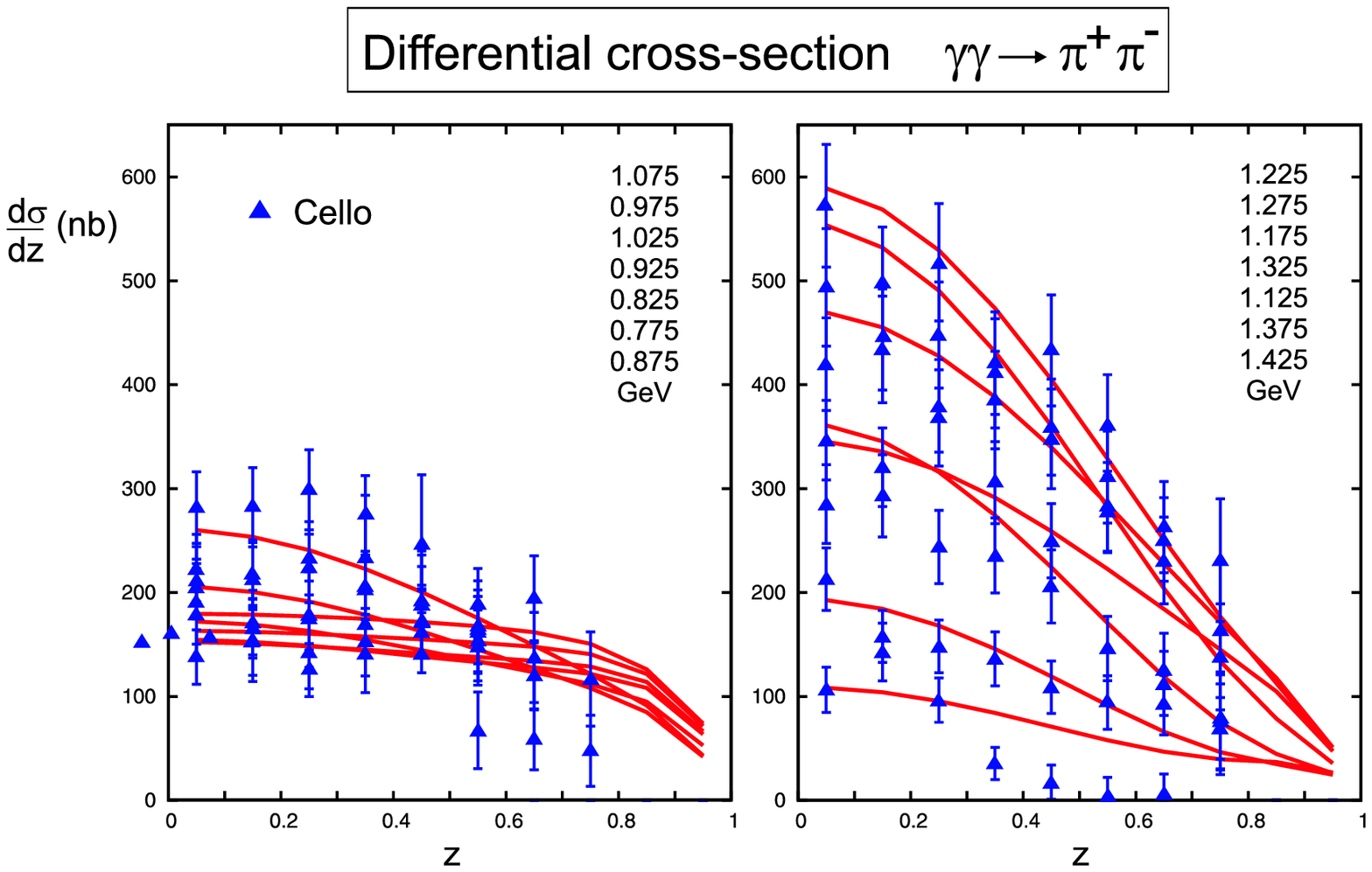,width=14.5cm}}
\vspace*{2pt}
\caption{\leftskip 1cm\rightskip 1cm{Differential cross-section for $\,\gamma\gamma\to\pi^+\pi^-\,$ from the CELLO experiment~\cite{harjes} compared with Solution A from the Amplitude Analysis. The numbers give the central energy in GeV of each angular distribution listed in order of the cross-section at $z=0$, where
$z\,=\,\cos\,\theta^*$. }}
\vspace{7mm}
\centerline{\psfig{file=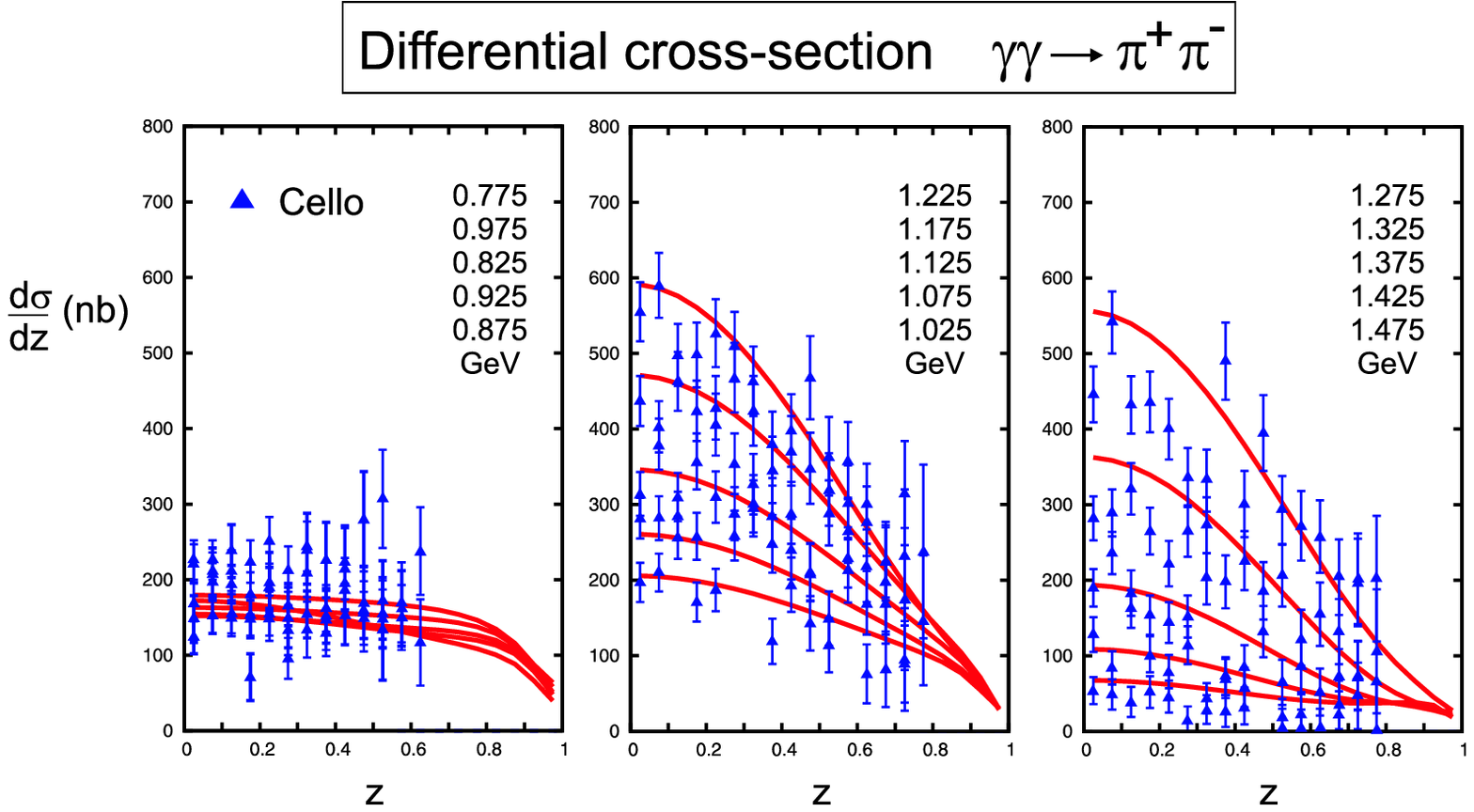,width=14.5cm}}
\vspace*{2pt}
\caption{\leftskip 1cm\rightskip 1cm{Differential cross-section for $\,\gamma\gamma\to\pi^+\pi^-\,$ from the CELLO experiment in finer binning in both energy and angle from those of Fig.~11~\cite{behrend}, compared with Solution A from the Amplitude Analysis. The numbers give the central energyin GeV of each angular distribution listed in order of the size of the cross-section at $z=0$, where
$z\,=\,\cos\,\theta^*$. }}
\end{figure}

\begin{figure}[p]
\centerline{\psfig{file=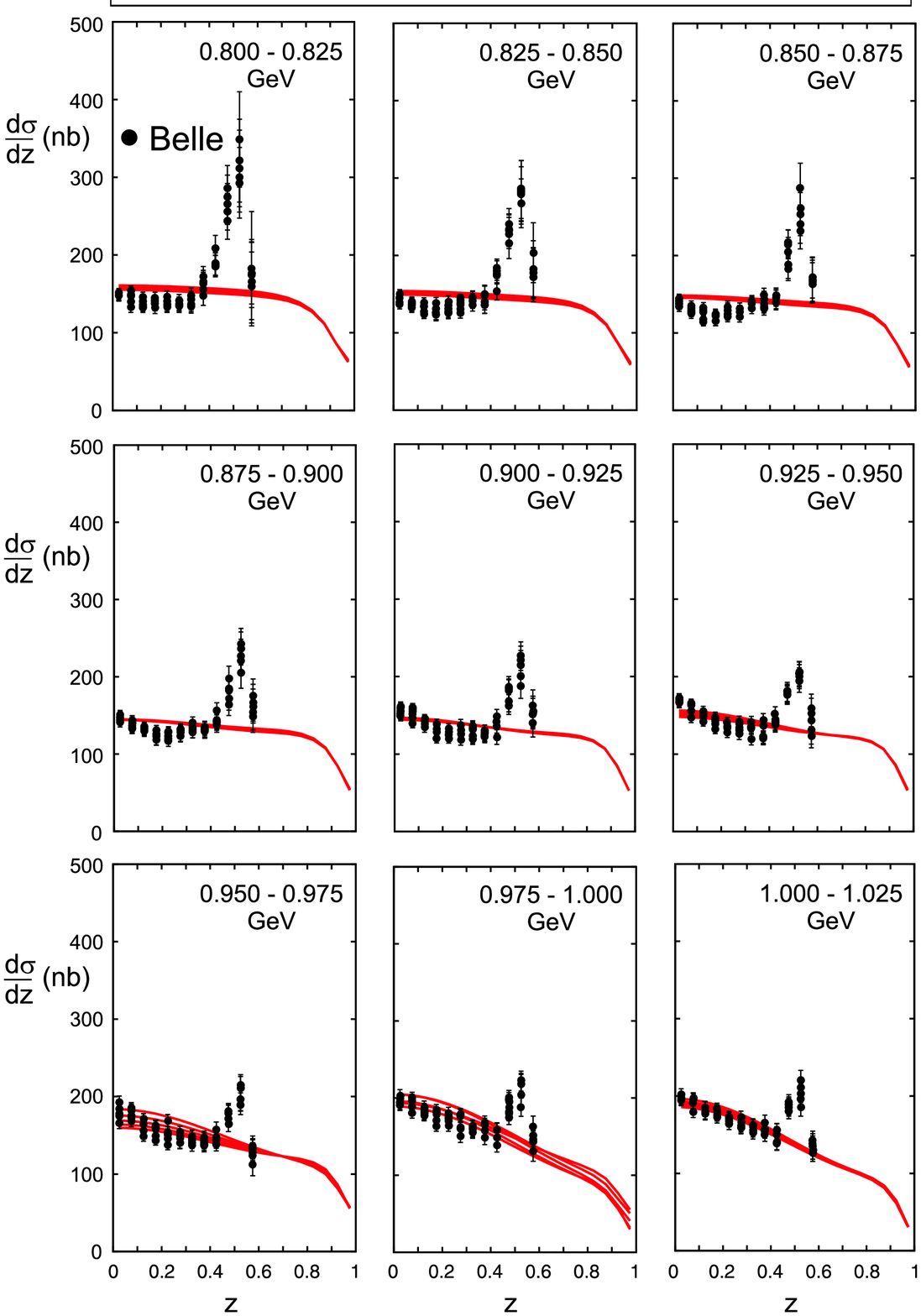,width=13.cm}}
\vspace*{2pt}
\caption{\leftskip 1cm\rightskip 1cm{Differential cross-section for $\,\gamma\gamma\to\pi^+\pi^-\,$ from the Belle experiment~\cite{abe} compared with Solution A from the Amplitude Analysis. Each graph  in 25 MeV intervals displays data and the result of the Amplitude Analysis in 5 MeV bins, so each graph has five sets of data and 5 curves.
$z\,=\,\cos\,\theta^*$.}}
\end{figure}

\begin{figure}[p]
\centerline{\psfig{file=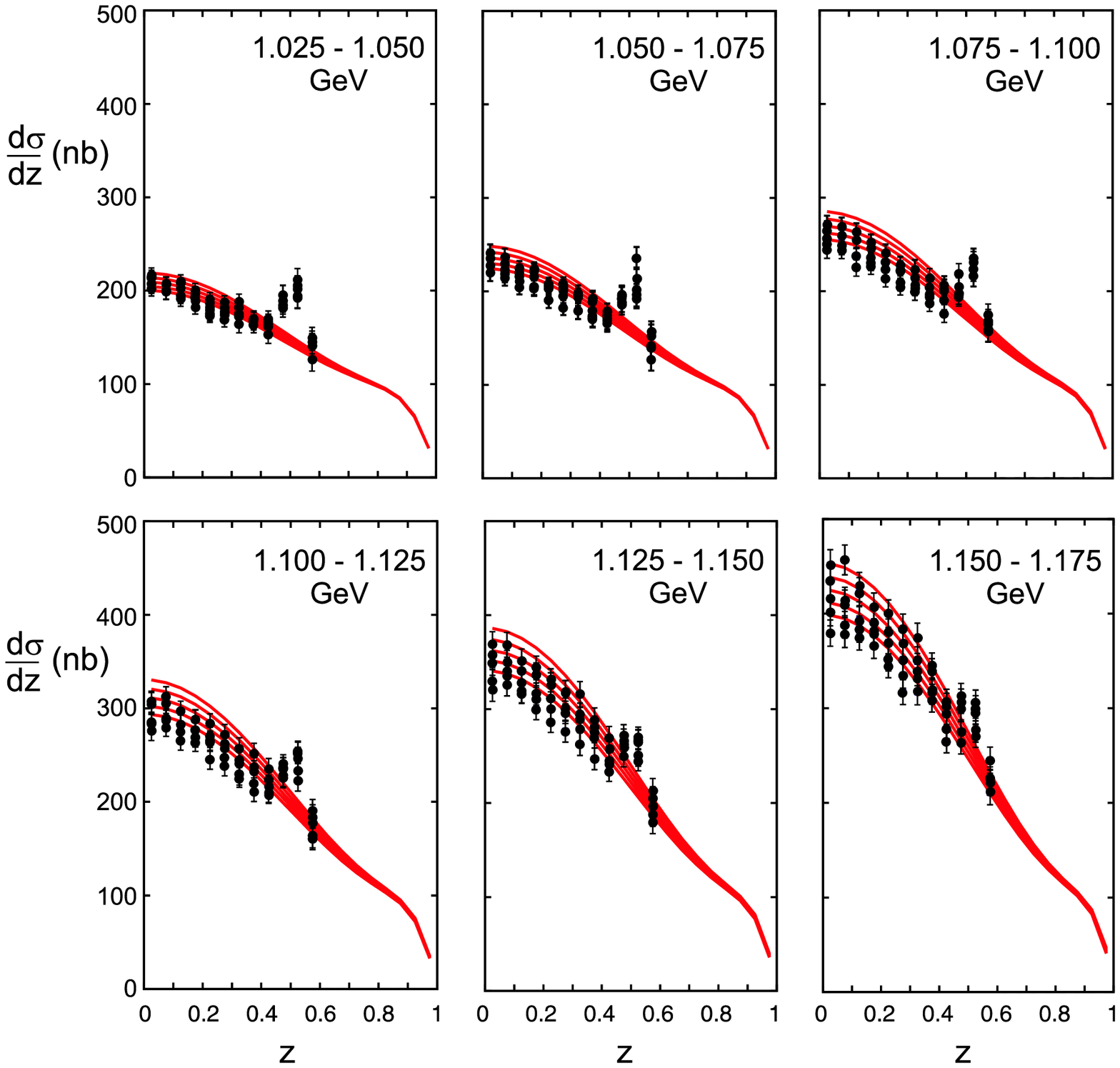,width=13.cm}}
\vspace*{2pt}
\caption{\leftskip 1cm\rightskip 1cm{Differential cross-section for $\,\gamma\gamma\to\pi^+\pi^-\,$ from the Belle experiment~\cite{abe} compared with Solution A from the Amplitude Analysis. Each graph  in 25 MeV intervals displays data and the result of the Amplitude Analysis in 5 MeV bins, so each graph has five sets of data and 5 curves.
$z\,=\,\cos\,\theta^*$.}}
\end{figure}

\begin{figure}[p]
\centerline{\psfig{file=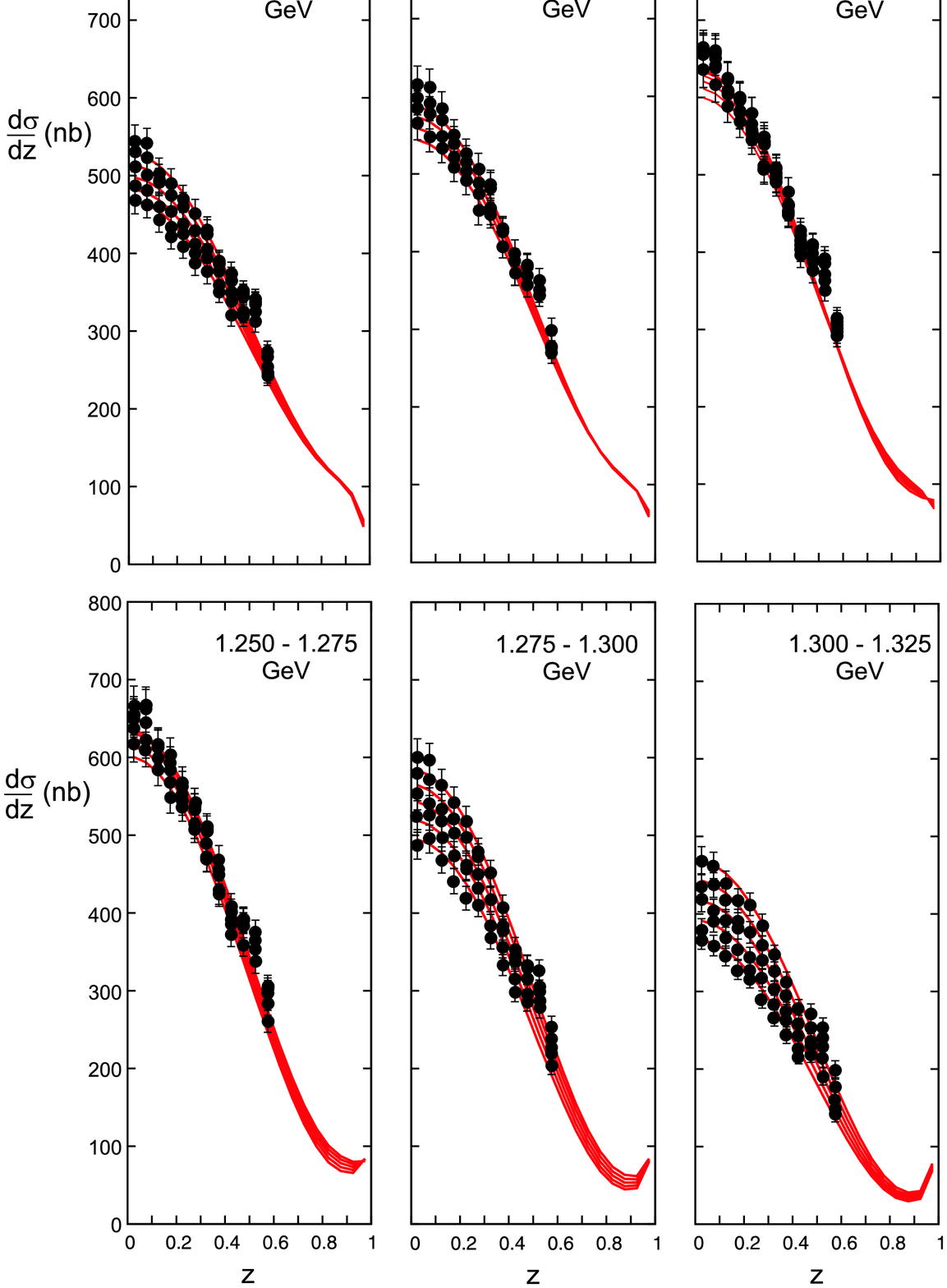,width=13.cm}}
\vspace*{2pt}
\caption{\leftskip 1cm\rightskip 1cm{Differential cross-section for $\,\gamma\gamma\to\pi^+\pi^-\,$ from the Belle experiment~\cite{abe} compared with Solution A from the Amplitude Analysis. Each graph  in 25 MeV intervals displays data and the result of the Amplitude Analysis in 5 MeV bins, so each graph has five sets of data and 5 curves.
$z\,=\,\cos\,\theta^*$.}}
\end{figure}

\begin{figure}[p]
\centerline{\psfig{file=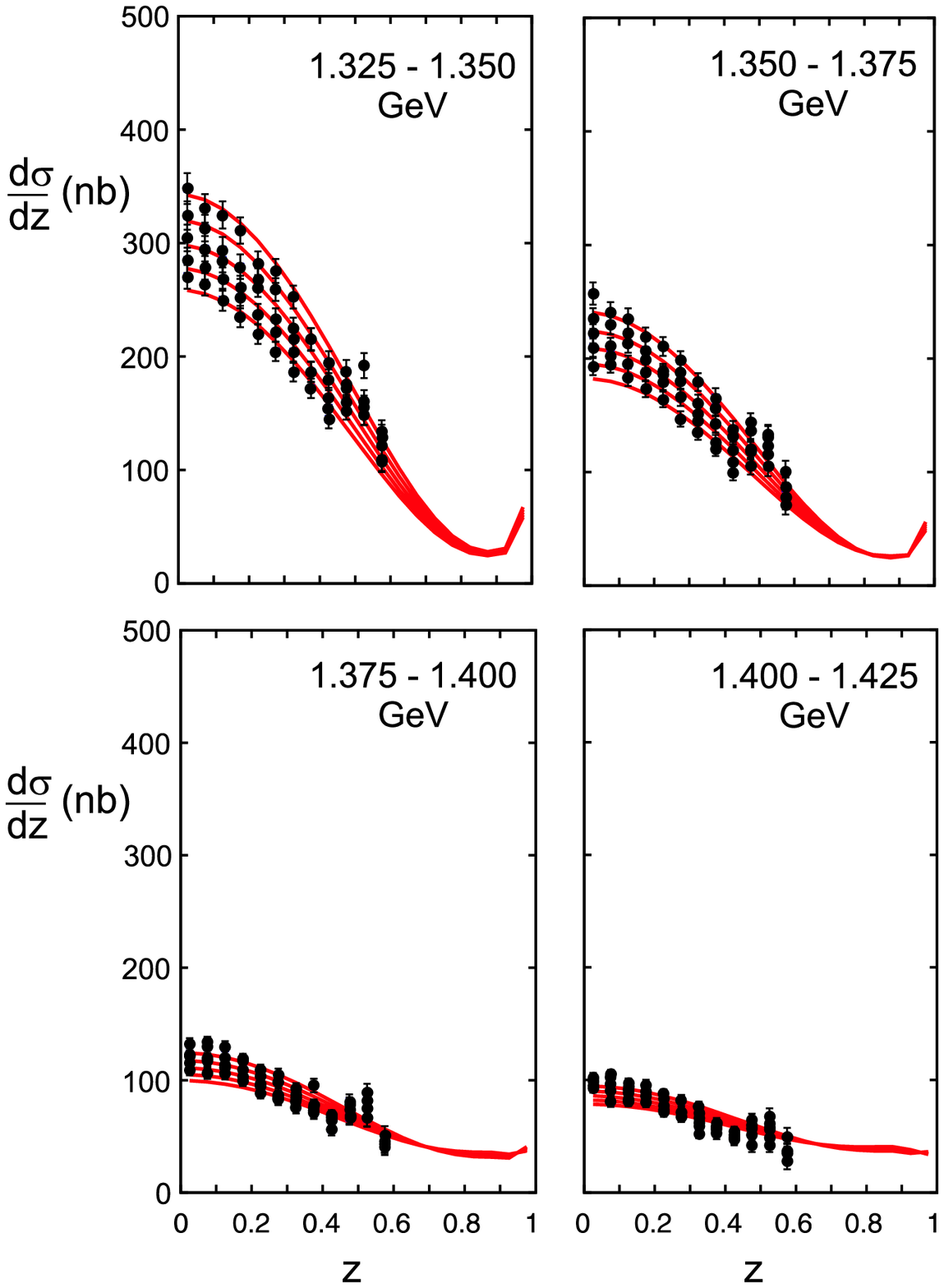,width=8.67cm}}
\vspace*{2pt}
\caption{\leftskip 1cm\rightskip 1cm{Differential cross-section for $\,\gamma\gamma\to\pi^+\pi^-\,$ from the Belle experiment~\cite{abe} compared with Solution A from the Amplitude Analysis. Each graph  in 25 MeV intervals displays data and the result of the Amplitude Analysis in 5 MeV bins, so each graph has five sets of data and 5 curves.
$z\,=\,\cos\,\theta^*$. }}
\end{figure}

\newpage
\newpage
We now turn to the angular distributions.
 In Figs.~11-18 we compare Solution A with the differential cross-sections from Mark II, CELLO and Belle for charged pions and Crystal Ball (1988 and 1992 results) for $\pi^0\pi^0$. While the huge dataset from Belle dominates, there appears to be a structure in the $|\cos \theta^*|\,\simeq\,0.45 - 0.55$ region below 1.1 GeV, seen in Fig.~15, to which we have just alluded. The fact that this possible \lq\lq measurement bias'' gives the fits a poor $\chi^2$ is one of the major reasons that  the Amplitude Analysis presented here cannot be more precise about the $f_0(980)$ radiative width. This poorly fitted \lq\lq structure'' in the angular distributions makes a whole range of solutions equally acceptable.
Solution B provides fits to the angular distributions that are not significantly different in the whole region below 1450 MeV, and so we do not show these fits separately. The somewhat worse $\chi^2$ apparent in Table 2 are barely discernible by eye.

\begin{figure}[p]
\centerline{\psfig{file=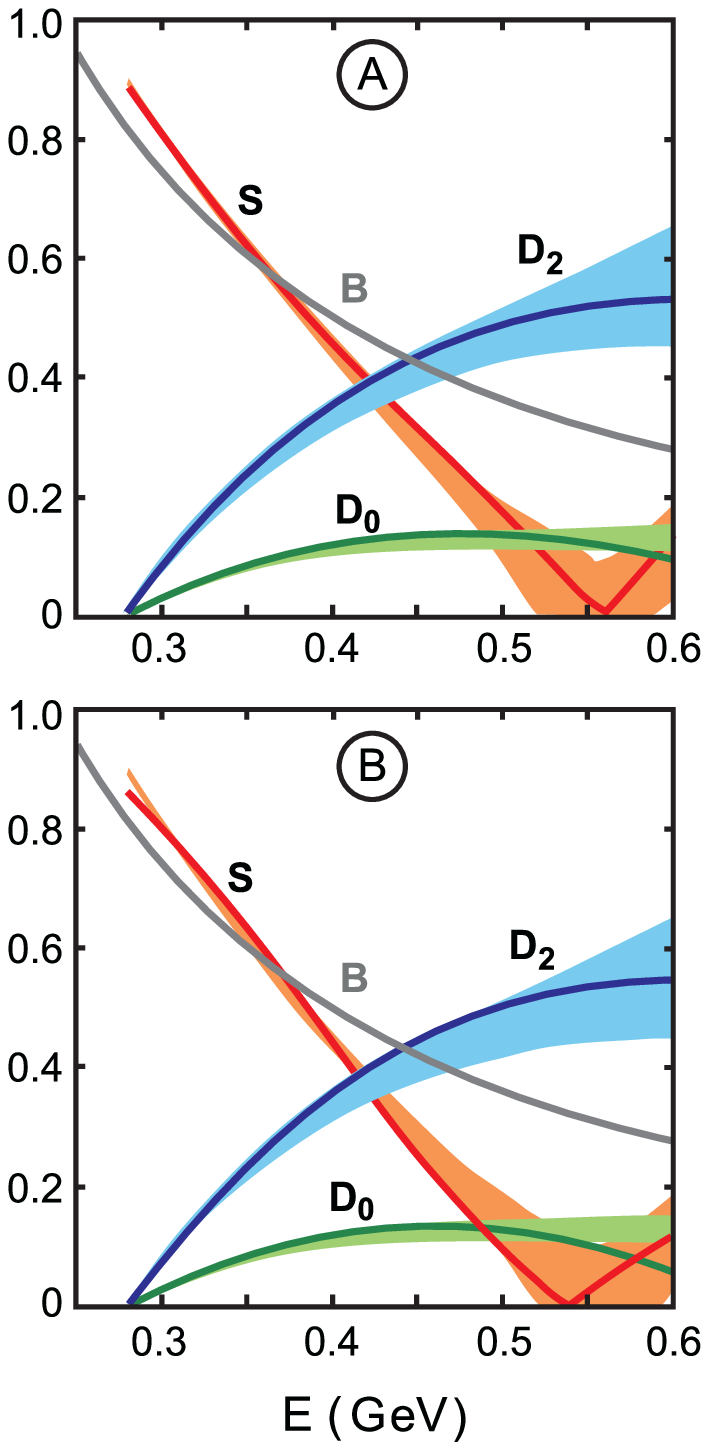,width=7.5cm}}
\vspace*{8mm}
\caption{\leftskip 1cm\rightskip 1cm{The bands display the moduli of the three dominant $I=0$ partial waves $S$, $D_0$ and $D_2$ at low energies as determined
by the dispersive analysis of the QED and chiral constraints discussed in the text. These amplitudes are the appropriate $F^I_{J\lambda}$ defined in Eqs.~(1-3). Precision close to threshold comes from the dominance
of one pion exchange to the cross-channel forces --- see Fig.~2. For orientation the $I=0$ Born $S$-wave (labelled $B$) is the black line. As the energy increases $\rho$, $\omega$, $a_1$, $\cdots$,
exchanges become increasingly important. Their contribution is less certain. This is reflected in the width of the bands. The solid lines are these same moduli
in the Solutions A and B.}}
\end{figure}

\begin{table}[p]
\caption{\leftskip 0.5cm\rightskip 0.5cm{Summary of contributions from each experiment to the total $\chi
^2$ for solution A and B. Here $\chi^2 _{\rm tot}$ is calculated by
dividing the sum of the  $\chi^2$'s for all datasets by the total number
of data-points we are fitting, namely 2267. $\chi^2 _{\rm average}$ is
computed by dividing the sum of the  $\chi^2$'s for each dataset
by the total number of data-points in that
experiment.  $^*$ indicates that the $\chi^2$ includes statistical and point-to-point systematic errors.
Only those with {\bf no} asterisk include  the overall
systematic error. }}
\vspace{-0.6cm}
\begin{center}
\begin{tabular}{||c|c||c|c||c|c||}
\hline \hline
\multicolumn{6}{||c||}{\rule[-0.4cm]{0cm}{10mm} favoured SOLUTION A
\parbox{1.2cm}{~~~~} $\chi ^2 _{{\rm tot}} = 2.88^*$ (0.91)} \\
\hline
\rule[-0.4cm]{0cm}{10mm} Experiment & Process & data-points & 
$\chi ^2_{{\rm average}}$ & $\chi^2_{\,{\rm Int. X-sect.}}$  & $\chi^2_{\,{\rm Ang. distrib.}}$ \\ 
\hline \hline
\rule[-0.8cm]{0cm}{18mm} 
Mark II & $\gamma \gamma \to \pi ^{+} \pi^{-}$ & 144 & 1.64$^*$ (1.02) &
 1.52$^*$ (0.94) & 1.80$^*$ (1.12) \\  
\hline 
\rule[-0.8cm]{0cm}{18mm}   
Cr. Ball & $\gamma \gamma \to \pi ^{0} \pi^{0}$ & 126 & 2.06$^*$ (1.72)  & 
2.04$^*$ (1.51) & 2.07$^*$ (1.81) \\  
\hline
\rule[-1.0cm]{0cm}{22mm}   
CELLO & $\gamma \gamma \to \pi ^{+} \pi^{-}$ & 333 & 1.76 & 
0.76 & 
\parbox{2.7cm}{1.57 \protect \\  from~Harjes \protect \\  2.10 \protect \\ from Behrend}\\
\hline
\rule[-0.8cm]{0cm}{18mm}
Belle & $\gamma\gamma\to\pi^+\pi^-$ & 1664 & 3.34$^*$ (0.67) & 0.85$^*$ (0.12) & 3.55$^*$ (0.72) \\
\hline \hline 
\end{tabular}
\vspace{0.2cm}

\begin{tabular}{||c|c||c|c||c|c||}
\hline \hline
\multicolumn{6}{||c||}{\rule[-0.4cm]{0cm}{10mm} SOLUTION B
\parbox{1.2cm}{~~~~} $\chi ^2 _{{\rm tot}} = 3.72^*$ (1.16)} \\
\hline
\rule[-0.4cm]{0cm}{10mm} Experiment & Process & data-points & 
$\chi ^2_{{\rm average}}$ & $\chi^2_{\,{\rm Int. X-sect.}}$  & $\chi^2_{\,{\rm Ang. distrib.}}$ \\ 
\hline \hline
\rule[-0.8cm]{0cm}{18mm} 
Mark II & $\gamma \gamma \to \pi ^{+} \pi^{-}$ & 144 & 2.01$^*$ (1.17) &
1.76$^*$ (1.06) & 2.32$^*$ (1.31) \\  
\hline 
\rule[-0.8cm]{0cm}{18mm}   
Cr. Ball & $\gamma \gamma \to \pi ^{0} \pi^{0}$ & 126 & 2.63$^*$ (2.27)  & 
1.95$^*$ (1.45) & 2.91$^*$ (2.60) \\  
\hline
\rule[-1.0cm]{0cm}{22mm}   
CELLO & $\gamma \gamma \to \pi ^{+} \pi^{-}$ & 333 & 2.10 & 
0.89 & 
\parbox{2.7cm}{1.94 \protect \\  from~Harjes \protect \\  2.22 \protect \\ from Behrend}\\
\hline
\rule[-0.8cm]{0cm}{18mm}
Belle & $\gamma\gamma\to\pi^+\pi^-$ & 1664 & 4.33$^*$ (0.89) & 1.00$^*$ (0.19) & 4.61$^*$ (0.95) \\
\hline \hline 
\end{tabular}
\end{center}
\end{table}

An important aspect of the Amplitude Analysis is that the individual partial waves are determined at low $\gamma\gamma$ energy by the QED low energy theorem and chiral constraints on the $\pi\pi$ final state interaction. These are implemented by the use of dispersion relations as discussed in Sect.~2 and as calculated in Refs.~\cite{morgam0,mpdaphne}. The precision at low $\pi\pi$ masses comes from the dominance of  one pion exchange to cross-channel forces (as in Fig.~2), modified by calculable final state interactions. All solutions are required to respect these properties. As the energy increases other crossed-channel exchanges become increasingly important. As their contribution is less certain, the
solutions are allowed to vary within the band of this uncertainty. In Fig.~19 
are shown  the bands for the $I=0$ $S, D_0, D_2$ waves and how Solutions A and B fit within these. The corresponding $I=2$ waves are similarly controlled by the meson-exchange amplitudes, but as they have much weaker final state interactions, they are closer to their Born amplitude. These are not shown. All waves with $J\ge 4$ are approximated by their Born contribution in the whole energy range fitted, {\it i.e.} up to 1.45 GeV. They generally make up only a small component of the $\pi^+\pi^-$ and $\pi^0\pi^0$ amplitudes and so this is a reasonable guide to their contribution in the energy range covered.

\newpage

\section{Two photon widths of the low mass isoscalars}

We are now in position to show the partial wave content of our Solutions A and B. In Fig.~20 we illustrate the contribution to the integrated cross-section of each of the $I=0$ and $I=2$ $S$, $D_0$, $D_2$ waves and $I=2$ $S$ and $D_2$ waves.
 Very close to threshold these are controlled by the one pion exchange Born term. If this were not modified by different final state interactions in the $I=0$ and $I=2$ channels the $I=0$ cross-sections would be exactly a factor of 2 larger than those with $I=2$.
One sees the difference is greater reflecting the stronger $\pi\pi$ interactions in the isoscalar channel. These bring about the marked dip in the $I=0$ $S$-wave seen around 550-600 MeV in Figs.~20,~21.
\begin{figure}[hb]
\vspace{3mm}
\centerline{\psfig{file=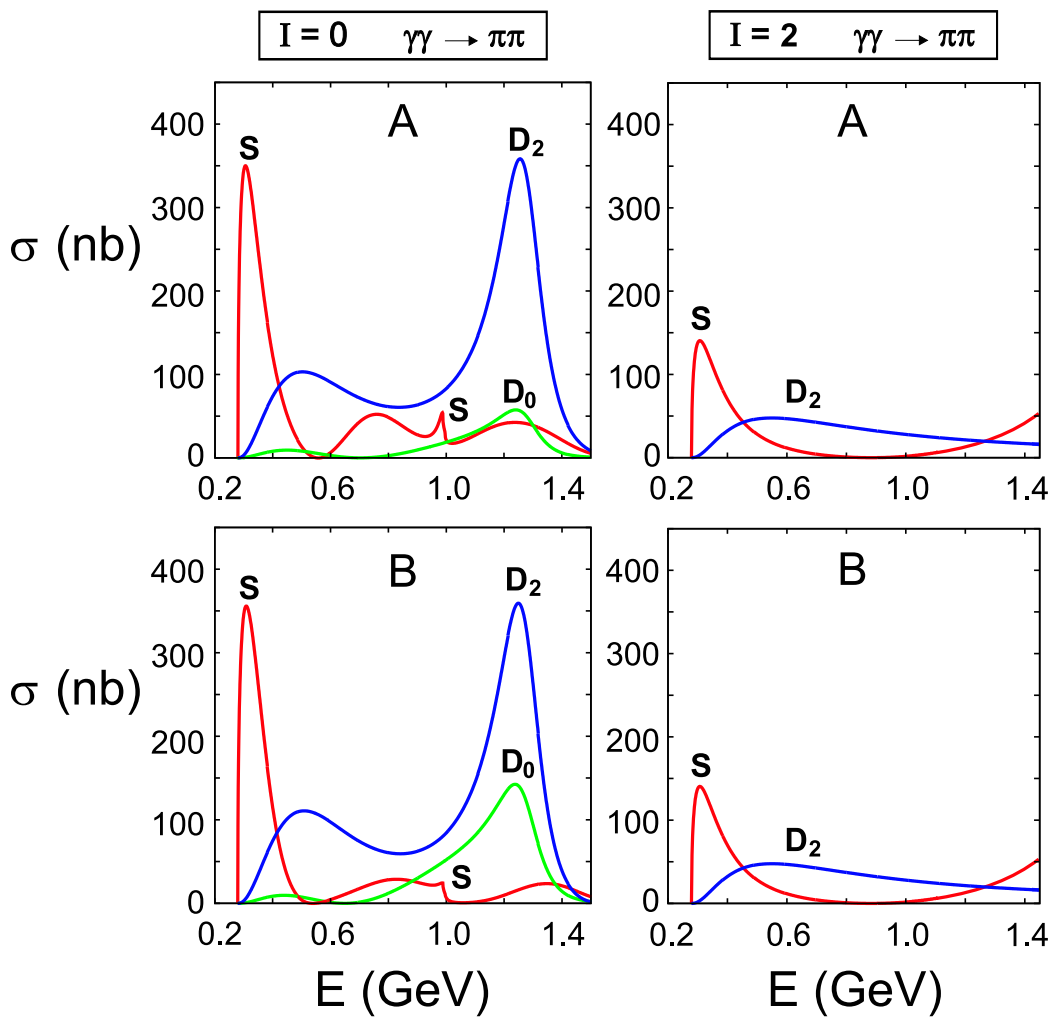,width=13.8cm}}
\vspace*{2pt}
\caption{\leftskip 1cm\rightskip 1cm{Contributions of the individual partial wave components to the full integrated cross-sections for Solutions A and B. Only the dominant waves are shown. The $I=2$ amplitudes are presumed to be computable  from the Born amplitude modified by known final state interactions, and so are the same for all solutions --- as for the $I=0$ amplitude close to threshold.}}
\end{figure}
 The sizeable difference between the Born amplitude and this $S$-wave is the signal of the $\sigma$-resonance, as noted in Ref.~\cite{mp-prl}. As the energy increases the isoscalar scalar component grows with a clear peak of the $f_0(980)$ in Solution A, and a little less so in Solution B. It then has a significant contribution  above 1200 MeV that one would associate with the $f_0(1370)$. The largest component above 500 MeV is, of course, the $f_2(1270)$. In the favoured Solution A this appears overwhelmingly in the $D_2$ wave, while in Solution B this is shared with the $D_0$ wave. To fit the data with limited angular information one has the obvious ambiguity of a significant $S$-wave in Solution A or a more complicated interference between $S$ and $D_0$ components in Solution B. Present data favour Solution A, but solutions like B are not yet ruled out.

\begin{figure}[p]
\centerline{\psfig{file=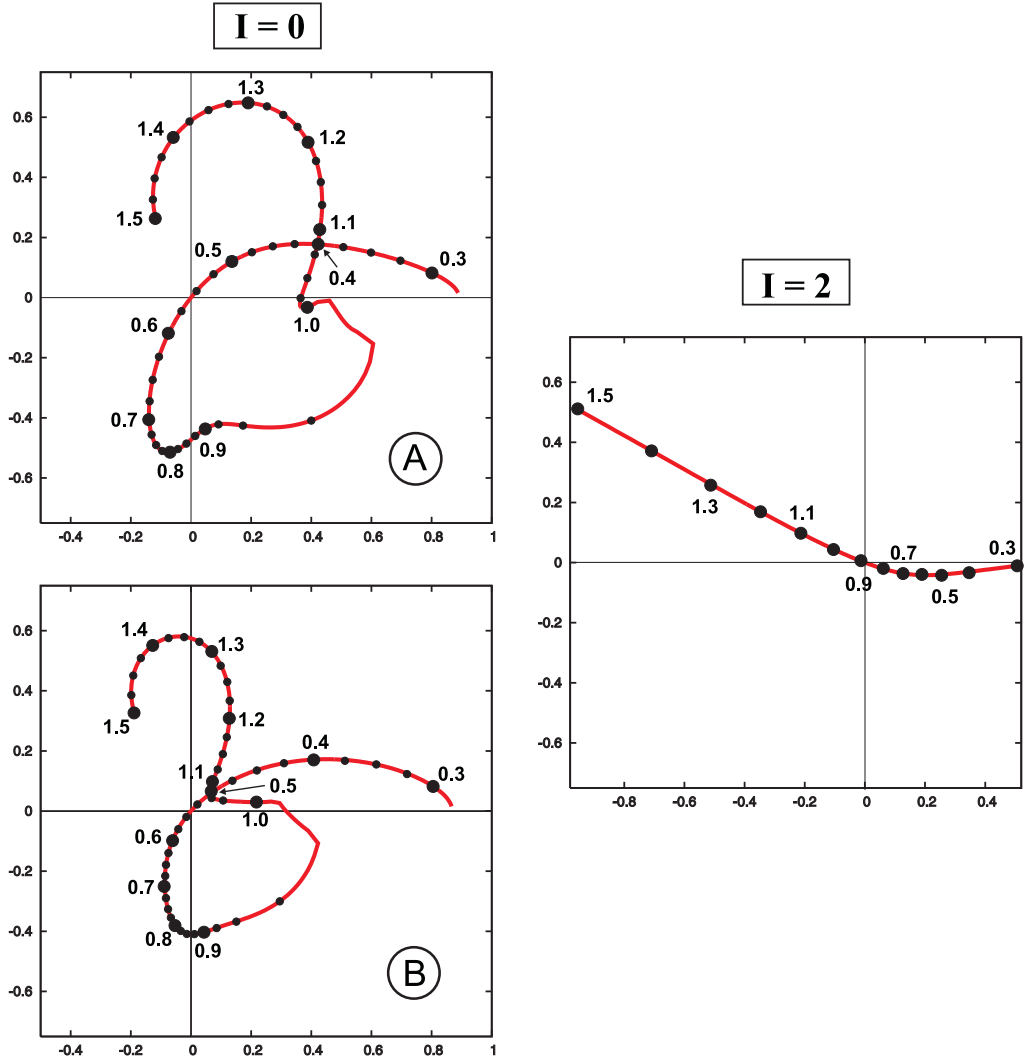,width=13.5cm}}
\vspace*{5mm}
\caption{\leftskip 1cm\rightskip 1cm{Argand plots for the $\gamma\gamma\to\pi\pi$ $I=0$ and $I=2$ $S$-wave amplitudes. The normalisation of these amplitudes, $F^{I=0}_{J=\lambda=0}$, is defined by Eqs.~(1-3). On the left are the $I=0$ $S$-wave for solutions A and B.
 The labels and bigger dots mark the energy every 0.1 GeV. The smaller dots are the intermediate energies every 25 MeV. The amplitudes A and B are very similar below 600 MeV, by construction, {\it cf.} Fig.~19. As seen the amplitudes move particularly fast between 950 and 1000 MeV because of the $f_0(980)$, that in A showing a somewhat bigger loop. The amplitudes display the expected \lq\lq kinks'' at $K^+K^-$ and $K^0{\overline K^0}$ thresholds.
On the right is the common $I=2$ $S$-wave with the dots labelling the energies every 0.1 GeV.}}
\end{figure}

In Fig.~21 are shown the Argand plots for the important $I=0$ $S$-wave amplitude for Solutions A and B. This $S$-wave amplitude encodes the effect in experiment of the $\sigma$, $f_0(980)$ and $f_0(1370)$ poles, the first and third of which are very wide. Consequently the amplitudes vary greatly from the real axis, where experiment is performed, to the pole positions. The Argand amplitudes are seen to move fastest between 950 and 1025 MeV, as expected from the far narrower $f_0(980)$. Solution A shows a larger movement indicating the larger radiative width for the $f_0(980)$ in this solution. Fig.~21 also shows the slowly changing
Argand plot for the $I=2$ $S$-wave amplitude, which is assumed to be the same for all solutions.

The unambiguous determination of the two photon width of each resonance requires the corresponding amplitudes to be continued to the pole on the nearby unphysical sheet of the energy plane. This is done by following the procedure set out
in Eqs. (42-49) in Ref.~\cite{boglione}. The residue of the relevant partial wave $\gamma\gamma\to\pi\pi$ amplitude at the resonance pole determines the product of couplings $g_\gamma g_\pi$ (or equivalently the functions ${\alpha_i}^I_{J\lambda}$ of Eqs.~(5,6) fix the ratio $g_\gamma/g_\pi$). The residue of the corresponding amplitude, ${\cal T}^I_J(\pi\pi\to\pi\pi)$ given here in Eq.~(5), then determines $g_\pi^2$. From these the two photon width is given by
\begin{equation}
\Gamma(R\to\gamma\gamma)\;=\; \frac {\alpha^2}{4 (2J+1) m_R}\,|g_\gamma|^2\quad,
\end{equation}
where $\alpha \simeq 1/137$ is the QED fine structure constant. 
The resulting two photon widths for the $\sigma$, $f_0(980)$ and $f_2(1270)$ are listed for Solutions A and B in Table 3. This is the primary output of this analysis. 

Our global fits determine the two photon width for the $f_0(980)$ to be 415 eV for the favoured solution A. However this lies in 
a range from
540 eV (for a solution qualitatively very close to A), down to 96 eV (as illustrated by Solution B). This is to be compared with various calculations dependent on the essential composition of this interesting state. A simple ${\overline s}s$ structure gives $\sim 200$ eV as found by Barnes~\cite{barnes}.  His calculation~\cite{barnesKK} of a ${\overline K}K$  composition
in the molecular model of Weinstein and Isgur~\cite{wi} gives $\sim 600$ eV.
 In contrast, more recently Hanhart {\it et al.}~\cite{hanhart} predict 220 eV in a ${\overline K}K$-molecular picture. This is comparable to the much earlier calculation of Achasov {\it et al.}~\cite{achasov}, who find, with a  ${\overline {qq}}qq$ composition, a prediction of
 $\sim 270$ eV, using Achasov's kaon loop model. Despite the claims of differing compositions, the essence of these latter two calculations is the same. 
If  the $f_0(980)$ width is indeed $\sim 415$ eV as in our favoured solution, this does not point unambiguously at any one particular modelling. However, as we have noted, solutions with smaller values of the $\gamma\gamma$ coupling are possible --- at present at a lower level of statistical probability, though nevertheless acceptable. Thus the 
Achasov {\it et al.}~\cite{achasov} and Hanhart {\it et al.}~\cite{hanhart}
 prediction may yet be realised.
 Clarification requires precision $\gamma\gamma\to\pi^0\pi^0$ results, as we discuss in detail below.

For the $f_2(1270)$ the helicity zero fraction is given --- though determined from the pole residue, it accords very closely with the ratio of cross-sections shown in Fig.~20. Folded into the error quoted on the $f_2(1270)$ is the overall  systematic uncertainty in each experiment. Though these are typically 10-12\%, the tension between the different datasets and their systematic errors means that a typical fit results in a smaller uncertainty on any given solution. 

The two photon width of the $\sigma$ is obtained by analytically continuing the $I=0$ $S$-wave amplitude, shown in Figs.~19,~20, to the $\sigma$-pole. The $\pi\pi$ amplitudes we use in this analysis have this pole located at the position found by Caprini {\it et al.}~\cite{caprini}, {\it viz.} at $E = 441 - i 272$ MeV.
The analytic continuation is performed using the twice subtracted dispersion relation  described in Ref.~\cite{mp-prl}, but including not just $\pi$, $\rho$, $\omega$-exchanges in the $t$ and $u$-channels (Figs.~4,~5), but also the $a_1$ and $b_1$, which the work of Ref.~\cite{oller} indicates are not insignificant.
This determines the ratio of the coupling $\sigma \to \gamma\gamma$ to that for $\sigma\to\pi\pi$, {\it i.e.} $g_\gamma/g_\pi$. The \lq\lq radiative width'' of Eq.~(7) requires the $\sigma\to\pi\pi$ coupling to be determined at the pole too. This value was estimated in Ref.~\cite{mp-prl} to be 0.55 GeV (see Eq.~(4) of Ref.~\cite{mp-prl} for the definition), while Leutwyler has given a smaller preliminary value of $(0.47 \pm 0.02)$ GeV~\cite{oller-leut}.  Using the latter value gives the results listed in Table~4, which are inevitably smaller than that found in Ref.~\cite{mp-prl} closer to those in Ref.~\cite{oller2}.

\begin{table}[t]
\caption{\leftskip 1.cm\rightskip 1.cm{Two-photon partial widths in keV corresponding to the two 
classes of solutions as determined from their pole residues. For each the fraction of the width provided by helicity zero is given: for the scalar resonances, this is, of course, 100\%. The error on the sigma's two photon width is largely due to the uncertainty in the $\sigma\to\pi\pi$ coupling at the pole, discussed in Refs.~\cite{oller,oller2} and in the text here.}}
\vspace{-1mm}
\begin{center}
\begin{tabular}{|l||c|c|c|}
\cline{2-4}
\multicolumn{1}{c|}{~~}
&\rule[-0.6cm]{0cm}{4mm}
 $f_2(1270)$ 
 &$f_0(600)/\sigma$ & $f_0(980)$ 
\rule[0.4cm]{0cm}{8mm}  \\ 
\hline 
&&&\\
Pole positions (GeV) & $1.276 - i0.094$ & $0.441 -i0.272$ &$1.001 - i0.016$\\
&&&\\
\hline \hline 
\rule[-0.4cm]{0cm}{8mm}
&&&\\[-4.mm]
$\Gamma(f_J\to\gamma\gamma)$: solution {\bf A}  & $3.14 \pm 0.20$ & $3.1 \pm 0.5$& ~~0.42~~ \\[-5mm]
~~~~~helicity zero fraction & 13\% & 100\% &100\%
\rule[0.4cm]{0cm}{8mm}   \\[2.5mm]
\hline
\rule[-0.4cm]{0cm}{8mm}
&&&\\[-4.mm] 
$\Gamma(f_J\to\gamma\gamma)$: solution {\bf B}  & $3.82 \pm 0.30$ & $ 2.4 \pm 0.4$ & 0.10  \\[-5mm]
~~~~~helicity zero fraction & 26\% & 100\% & 100\%   
\rule[0.4cm]{0cm}{8mm}   \\[2.5mm]
\hline 
\end{tabular}
\vspace{-2mm}
\end{center}
\end{table}

\begin{figure}[p]
\centerline{\psfig{file=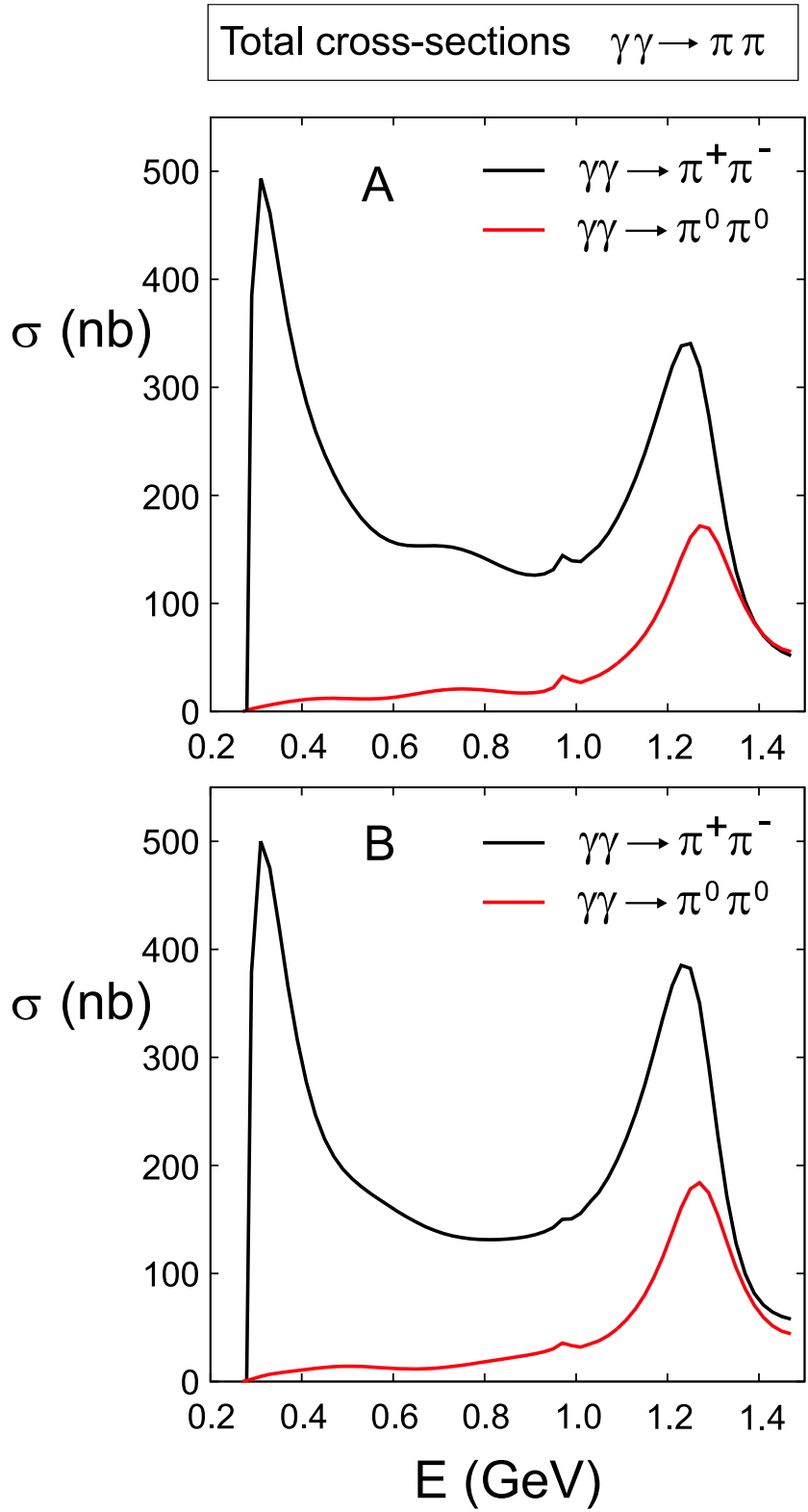,width=11.cm}}
\vspace*{2pt}
\caption{\leftskip 1cm\rightskip 1cm{Total cross-sections for \ggppch and $\pi^0\pi^0$ for Solutions A and B evaluated in 20 MeV energy bins with the bin centres connected by continuous lines.}}
\end{figure}

\begin{figure}[p]
\centerline{\psfig{file=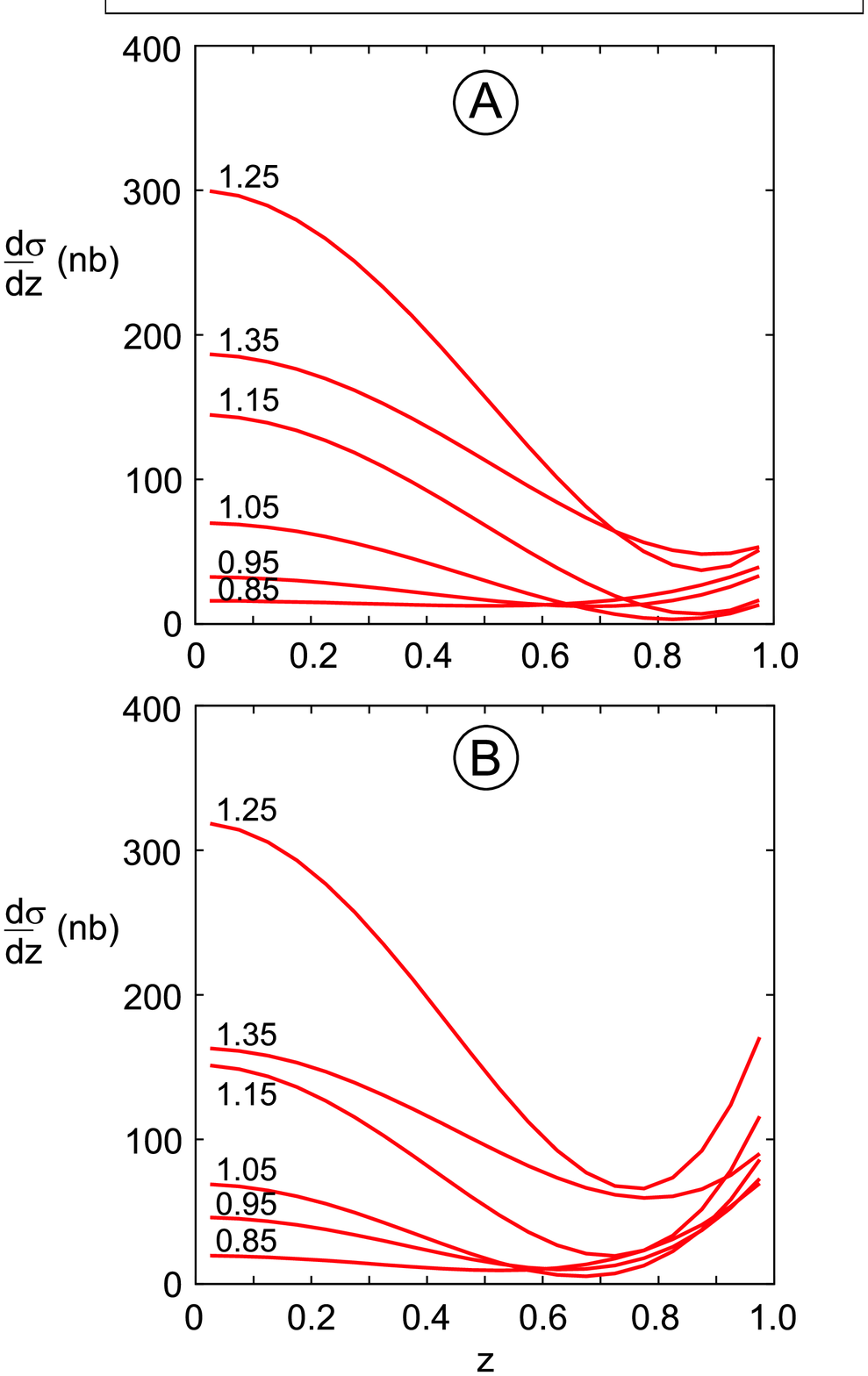,width=11.cm}}
\vspace*{2pt}
\caption{\leftskip 1cm\rightskip 1cm{Differential cross-section for \ggppn for Solutions A and B evaluated in 20 MeV energy bins centred on the energies specified in GeV. Even if the forthcoming precision results from Belle cover only $|\,z\,=\,\cos \theta^*\,|\,\le\, 0.8$, it should be possible to differentiate between these solutions.}}
\end{figure}

In Fig.~22 are shown the total cross-sections for $\gamma\gamma\to\pi^+\pi^-$ and $\pi^0\pi^0$ for both Solutions A and B evaluated in 20 MeV bins. If the $f_0(980)$ radiative width is \lq\lq large'' as in Solution A, with a width of 415 eV, then the peak in the 950 MeV region becomes even more pronounced in both the charged and neutral pion data if the full angular range were accessible. In
contrast, a much smaller $I=0$ $S$-wave component is compatible with the Belle charged pion data over its limited angular range, if the peak near 950 MeV is produced by sizeable $S-D_0$ wave interference. Such interferences disappear over the full angular range when the partial wave components become orthogonal. This explains the difference in cross-sections between Solutions A and B in Fig.~20,
where the contribution from individual partial waves with $I=0$ and $2$ are shown.

Precise measurement of the $\gamma\gamma\to\pi^0\pi^0$ channel can distinguish between these two classes of solution. Not only are the total cross-sections different as displayed in Fig.~22, but importantly the angular distributions shown in Fig.~23 are quite distinct particularly at large $z = \cos\theta^*$. With data promised from Belle in the near future, this ambiguity may well be resolved.

\section{Discussion}

An essential prerequisite of a meaningful Amplitude Analysis of 
the process \ggpp is differential cross-section data covering
the widest possible angular range for both the $\pi^+\pi^-$ and $\pi^0\pi^0$ channels. While the Belle collaboration provides data of enormous statistics
on the charged pion channel, this only covers 60\% of the the range in
$\cos\theta^*$. To make an Amplitude Analysis feasible, the data on the
neutral pion channel from Crystal Barrel~\cite{cb88,cb92}, covering
70-80\% of the angular range, with limited statistics (see Table~1, and Figs.~7,~11) are
given sizeable weight, so that they play a role of equal importance to the several datasets on the $\pi^+\pi^-$ mode. The resulting solutions are of good quality, as indicated in Table~2 and Figs.~7-18, following all datasets adequately.
While Solution~A with a relatively strong $f_0(980)$ signal corresponding to
a two photon width of 415 eV is preferred in terms of $\chi^2$ (Table~2),
partial wave solutions with a range of $f_0(980)$ components are possible.
As illustrated by Solution B, a two photon width of the $f_0(980)$ as 
small as 100 eV is at the extreme of acceptability. This is made possible
because the limited angular coverage allows the data to be described by a smaller $S$-wave, but larger $D_0$ wave, as shown in Fig.~20. 
At the other extreme a solution, with the qualitative features of Solution A but much higher $\chi^2$, has a radiative width for the $f_0(980)$ of 540 eV.
Precision results
on $\gamma\gamma\to\pi^0\pi^0$ should resolve this remaining ambiguity, as illustrated in Figs.~22,~23.

Achasov and Shestakov~\cite{Achasov:2007qr} have analysed the Belle integrated $\pi^+\pi^-$ cross-section in terms of model amplitudes in which key resonances
have both a direct and meson loop contributions. While they find adequate fits
to the charged pion data, their amplitudes fail to reproduce the $\pi^0\pi^0$
 cross-section of Crystal Ball above 800 MeV. In contrast, our Amplitude Analysis treats both charged and neutral channels simultaneously and fits adequately
 not just the integrated cross-sections,
but the angular distributions as well.

In the simplified analysis of Ref.~\cite{Achasov:2007qr}, the Belle data lead to a somewhat 
larger
$f_2(1275)\to\gamma\gamma$ width than previously found. This is at first sight not surprising looking at the comparison of experimental data shown in Fig.~2. The PDG quotes~\cite{pdg} $\Gamma(f_2(1275)\to\gamma\gamma)\,=\,2.60 \pm 0.24$ keV
largely based on the earlier (pre-Belle) Amplitude Analysis of Ref.~\cite{boglione}, which gives $2.84 \pm 0.35$ keV.
Our presently favoured solution A, in which the $f_2(1275)$ is produced predominantly with helicity two,  gives a comparable result of $\Gamma(f_2(1275)\to\gamma\gamma) =3.14 \pm 0.20  $ keV. In contrast, Solution B  with its large $D_0$ component, the $f_2(1275)$ width rises to $3.82 \pm 0.30  $ keV. Once again forthcoming $\gamma\gamma\to\pi^0\pi^0$ data have the power to reduce these uncertainties dramatically. We eagerly await the completion of the analysis of these data. Their inclusion in an Amplitude Analysis will hopefully lead to a consistent set of two photon couplings for the low mass isoscalar states: $f_0(980)$ and $f_2(1270)$. Two photon couplings are a key window on the detailed structure of these
low mass isoscalar states. Results from Belle make definitive conclusions for these within reach.   To reduce still further the uncertainties in the $\gamma\gamma$ width of the $\sigma$ requires precision charged and neutral pion data between threshold and 700 MeV~\cite{mpreviews}. The introduction of appropriate taggers
in an upgraded DA$\Phi$NE machine at Frascati~\cite{daphne2} may well make this feasible too.
We may then be certain which of the possible compositions~\cite{mpreviews,klempt,mp-menu} these scalars have.

\section*{Acknowledgments}
 
MRP acknowledges partial support of the EU-RTN Programme, Contract No. MRTN--CT-2006-035482, \lq\lq Flavianet'' for this work. He is also grateful to both
Tokyo Institute of Technology and KEK for travel support, which enabled this collaboration. 
TM, SU and YW acknowledge support from the Japan Society of the Promotion of Science.


\end{document}